\documentclass[12pt, oneside, onecolumn, floatfix]{article}

\usepackage{epsfig}
\usepackage[a4paper]{geometry}
\usepackage{enumitem}

\usepackage{jcappub}

\usepackage{caption}
\usepackage{amsmath}
\usepackage{amssymb}
\usepackage{graphicx}

\usepackage{booktabs}
\usepackage{epstopdf}

\usepackage{gensymb}
\usepackage{subfigure}

\usepackage{multirow}

\usepackage[colorlinks=true,citecolor=blue,linkcolor=blue]{hyperref}
\usepackage[capitalise,noabbrev]{cleveref}

\usepackage{pifont}
\newcommand{\cmark}{{\color{mediumseagreen} \ding{51}}}%
\newcommand{\xmark}{{\color{red} \ding{55}}}%

\usepackage{tikz,xcolor,hyperref}

\definecolor{lime}{HTML}{A6CE39}
\DeclareRobustCommand{\orcidicon}{\hspace{-3mm}
	\begin{tikzpicture}
		\draw[lime, fill=lime] (0,0) 
		circle [radius=0.16] 
		node[white] {\hspace{0.1mm}{\fontfamily{qag}\selectfont \tiny ID}};
		\draw[white, fill=white] (-0.07,0.1) 
		circle [radius=0.01];
	\end{tikzpicture}
	\hspace{-5mm}
}

\foreach \x in {A, ..., Z}{\expandafter\xdef\csname 
orcid\x\endcsname{\noexpand\href{https://orcid.org/\csname orcidauthor\x\endcsname}
		{\noexpand\orcidicon}}
}

\definecolor{viridian}{rgb}{0.25, 0.51, 0.43}
\definecolor{mediumseagreen}{rgb}{0.24, 0.7, 0.44}
\definecolor{otterbrown}{rgb}{0.4, 0.26, 0.13}
\definecolor{saddlebrown}{rgb}{0.55, 0.27, 0.07}
\definecolor{americanrose}{rgb}{1.0, 0.01, 0.24}
\definecolor{ao}{rgb}{0.0, 0.0, 1.0}

\newcommand\myshade{80}
\colorlet{mylinkcolor}{violet}
\colorlet{mycitecolor}{red}
\colorlet{myurlcolor}{ao}

\hypersetup{
	linkcolor  = mylinkcolor!\myshade!black,
	citecolor  = mycitecolor!\myshade!black,
	urlcolor   = myurlcolor!\myshade!black,
	colorlinks = true
}

\title{Neutrino constraints on inelastic dark matter captured in the Sun}

\author[a\hspace{-5pt}]{Bhavesh Chauhan\orcidA{}}
\emailAdd{bhavesh-chauhan@uiowa.edu}

\author[a\hspace{-5pt}]{, Mary Hall Reno\orcidB{}}
\emailAdd{mary-hall-reno@uiowa.edu}
    
\author[b,c]{,\\Carsten Rott\orcidC{}}
\emailAdd{rott@physics.utah.edu}

\author[d,e]{, Ina Sarcevic\orcidD{}}
\emailAdd{ina@physics.arizona.edu}

\affiliation[a\hspace{1pt}]{Department of Physics and Astronomy, University of Iowa, \\Iowa City. IA 
52242, USA}

\affiliation[b\hspace{1pt}]{Department of Physics and Astronomy, University of Utah, \\Salt Lake City, 
UT 
84112, USA}

\affiliation[c\hspace{1pt}]{Department of Physics, Sungkyunkwan University, \\ Suwon, Rep. of Korea}

\affiliation[d\hspace{1pt}]{Department of Physics, University of Arizona, \\Tucson, AZ 85721, USA}
\affiliation[e\hspace{1pt}]{
Department of Astronomy and Steward Observatory,
University of Arizona, \\Tucson, AZ 85721,USA}

\abstract{
The flux of neutrinos from annihilation of gravitationally captured dark matter in the Sun has significant constraints from direct-detection experiments. However, these constraints are relaxed for inelastic dark matter as inelastic dark matter interactions generate less energetic nuclear recoils compared to elastic dark matter interactions. In this paper, we explore the possibility for large volume underground neutrino experiments to detect the neutrino flux from captured inelastic dark matter in the Sun. The neutrino spectrum has two components: a mono-energetic ``spike" from pion and kaon decays at rest and a broad-spectrum ``shoulder" from prompt primary meson decays. We focus on detecting the shoulder neutrinos from annihilation of hadrophilic inelastic dark matter with masses in the range 4--100 GeV and the mass splittings in up to 300 keV. We determine the event selection criterion for DUNE to identify GeV-scale muon neutrinos and anti-neutrinos originating from hadrophilic dark matter annihilation in the Sun, and forecast the sensitivity from contained events. We also map the current bounds from Super-Kamiokande and IceCube on elastic dark matter, as well as the projected limits from Hyper-Kamiokande, to the parameter space of inelastic dark matter. We find that there is a region of parameter space that these neutrino experiments are more sensitive to than the direct-detection experiments. For dark matter annihilation to heavy-quarks, the projected sensitivity of DUNE is weaker than current (future) Super (Hyper) Kamiokande  experiments. However, for the light-quark channel, only the spike is observable and DUNE will be the most sensitive experiment.     
}

\def\C3F8{$\rm C_3 F_8$}

\begin{document}
	
\maketitle
	
\section{Introduction}

There is overwhelming evidence for the existence of dark matter (DM) in our universe. The most widely accepted model of cosmology, $\Lambda$CDM, points towards a non-baryonic and non-relativistic species that constitutes nearly 25\% of the energy budget of our universe \cite{Planck:2018vyg}. However, the particle nature of DM, i.e., its mass, spin, and interactions, is yet unknown. As the Earth and Sun move through the halo of DM trapped in our galaxy, feeble interactions between DM and visible matter are expected. Direct detection experiments on Earth aim to detect the nuclear and electronic recoils produced in these scatterings. On the other hand, DM scattering off nuclei in Sun can lead to gravitational capture and DM is gradually accumulated in the Sun \cite{Press:1985ug,Griest:1986yu,Gould:1987ju,Gould:1991hx}. The captured DM can annihilate or decay into Standard Model particles, producing a flux of neutrinos that can be detected by large volume neutrino detection experiments on Earth such as Super-Kamiokande \cite{Super-Kamiokande:2004pou,Super-Kamiokande:2011wjy, Super-Kamiokande:2015xms}, ANTARES \cite{ANTARES:2013vvr, ANTARES:2016xuh}, and IceCube \cite{IceCube:2012ugg, IceCube:2021xzo}. 

The main goal of this paper is to take a detailed look at the sensitivity of the Deep Underground Neutrino Experiment (DUNE) \cite{DUNE:2020lwj, DUNE:2020ypp} to the flux of neutrinos from DM annihilation in the Sun, and to compare with limits from other neutrino and direct-detection experiments. This topic has been extensively studied in the context of elastic scattering of DM with the conclusion that most of the parameter space that can be probed by neutrino experiments is already ruled out by direct-detection experiments, except perhaps Hyper-Kamiokande \cite{Hyper-Kamiokande:2018ofw} can do marginally better in some channels \cite{Bell:2021esh}. As a result, we focus our attention on inelastic dark matter proposed in Ref. \cite{Tucker-Smith:2001myb} to address the discrepancy between the observed annual modulation in DAMA \cite{DAMA:2000mdu, DAMA:2008jlt, DAMA:2010gpn} and the null results in other direct-detection experiments. We assume a simplified model where the dark sector comprises two species, $\chi_1$ and $\chi_2$, with an endothermic mass splitting, $\delta = M_2 - M_1 > 0$, and contact interaction with the quarks. In this work, we only study hadrophilic interactions and the results for leptophilic interactions can be analogously obtained \cite{Garani:2017jcj}.

As we discuss below, the large volume underground neutrino experiments can probe novel regions of the parameter space for inelastic dark matter that are inaccessible to direct-detection experiments. This is owning to the fact that the dark matter capture in the Sun is less affected by the mass splitting than direct detection experiments (see \Cref{sec:idm}). The subsequent neutrino flux from inelastic DM annihilation in the Sun can be detected by the neutrino experiments \cite{Shu:2010ta}. The fraction of atmospheric neutrinos that is coincident with the position of the Sun constitute the main background. The main challenge to the sensitivity of the neutrino experiments is the poor source-pointing resolution at these energies, which can be attributed to the lack of one-to-one correspondence between the direction of incident neutrino and detected charged lepton.

It has been pointed out in Refs. \cite{Rott:2012qb, Bernal:2012qh} that dark matter annihilation to quarks in the Sun results in a large flux of mono-energetic neutrinos from pions and kaons that decay-at-rest after thermalization (called $\pi$DAR and KDAR respectively). The neutrinos from $\pi$DAR have energies $\sim30$ MeV and are extremely difficult to detect due to large atmospheric neutrino background. As a result, the community has focused on detection of the neutrinos from KDAR that have energies $\sim236$ MeV that allow for approximate reconstruction of the incident neutrino direction \cite{Rott:2015nma, Rott:2016mzs, Rott:2019stu, DUNE:2021gbm}. This mono-energetic feature in the neutrino spectrum is often called a ``hump'' or ``spike". While a positive detection of these spike neutrinos will be a smoking-gun signal of DM capture in the Sun, the lack of any spectral information will result in a degeneracy between the reconstructed dark matter parameters such as its mass, cross section, and the annihilation channel. The observation of spike neutrinos in DUNE gives a poor understanding of the particle nature of the dark matter, and a complementary probe would be beneficial. With this motivation, we focus on the other part of the emitted neutrino spectrum called the ``shoulder" which comprises neutrinos from the decay of primary mesons prior to thermalization in the Sun. The shoulder is a broad spectrum feature that extends up to the mass of the annihilating dark matter, and strongly depends on the annihilation channel. We refer the reader to \Cref{fig:sample} for an example of the neutrino spectrum from annihilation of 25 GeV dark matter in the Sun.

\begin{table}[t]
\renewcommand{\arraystretch}{1.2}
\centering
\begin{tabular}{c c c c c c c c}
\toprule
\multicolumn{2}{c }{} & \multicolumn{3}{c}{$\nu_e$ } & \multicolumn{3}{c }{$\nu_\mu$ } \\
\multicolumn{2}{c }{} & $E_\nu$& $\theta_\nu$ & Ref. & $E_\nu$ & $\theta_\nu$ & Ref.\\ \hline
\multirow{3}{*}{\shortstack{Spike \\ (\,QEL\,)}} & SuperK & \cmark & \xmark & \cite{Rott:2015nma, Rott:2016mzs} & \cmark & \xmark & -- \\
& IceCube &\xmark & \xmark & -- &\xmark & \xmark & --\\ 
& DUNE & \cmark & \cmark & \cite{Rott:2015nma, Rott:2016mzs} & \cmark & \cmark & \cite{Rott:2016mzs},\cite{DUNE:2021gbm}*\\ \midrule
\multirow{3}{*}{\shortstack{Shoulder \\ (\,DIS\,)}} & SuperK & \cmark & \cmark & \cite{Super-Kamiokande:2015xms}*$\,^{\#}$ & \cmark & \cmark & \cite{Super-Kamiokande:2004pou, Super-Kamiokande:2011wjy, Super-Kamiokande:2015xms}*  \\
& IceCube &\cmark & \xmark & -- &\cmark & \cmark & \cite{IceCube:2012ugg, IceCube:2021xzo}* \\
& DUNE&\cmark & \cmark & \cite{Bueno:2004dv} &\cmark & \cmark& This Work \\ 
\bottomrule
\end{tabular}
\caption{\label{tab:bird} We provide a bird's eye view of the capabilities of the neutrino experiments to detect the spike and shoulder neutrinos from the dark matter annihilation in the Sun. The spike neutrinos undergo quasi-elastic like (QEL) scattering and the shoulder neutrinos mostly undergo deep-inelastic scattering (DIS). We show the ability (inability) of the detectors to \emph{reasonably} reconstruct the incident neutrino energy and direction with a \,\cmark\, (\,\xmark\,). Note that for the shoulder, sensitivity is same for neutrinos and anti-neutrinos. We also provide references (Ref.) for the corresponding study, and mention that this work discusses detection of the shoulder muon neutrinos in DUNE. The references marked with * are by the respective collaboration. $\,^{\#}$Events from all neutrino flavors are included in Ref. \cite{Super-Kamiokande:2015xms}. See text for details.}
\end{table}

Observation of shoulder neutrinos will help to understand the flavor structure of dark matter interactions with quarks. If dark matter annihilates to light quarks ($u\bar{u}$, $d\bar{d}$, $s\bar{s}$), then the KDAR neutrinos are the only detectable component. The KDAR neutrinos produced in the Sun are muon neutrinos but the flavor conversion between the production point and the detector results in all flavors of these mono-energetic neutrinos in the detector. In a water Cherenkov detector like Super-Kamiokande, only the $\nu_e$ component is detectable as $\nu_\mu$ is very close to the detector threshold. DUNE can utilize the excellent low-energy resolution of Liquid Argon Time Projection Chamber (LArTPC) and detect both $\nu_e$ and $\nu_\mu$ components of the spike. In Ref. \cite{Rott:2015nma}, the sensitivity of Super-Kamiokande and DUNE to the $\nu_e$-spike has been studied. However, the source-pointing resolution of a water Cherenkov detector from quasi-elastic scattering needs to be examined carefully, and we look forward to the detailed study by the Super-Kamiokande collaboration in future. The sensitivity of DUNE to the $\nu_\mu$-spike events is studied in Ref. \cite{DUNE:2021gbm}, and limits are provided for inelastic dark matter assuming a mass splitting of 50 keV. In this work, we include a wider parameter space, and we discuss comparisons with direct-detection experiments.

If dark matter annihilates to heavy quarks (such as $b\bar{b}$ and/or $c\bar{c}$), then the shoulder neutrinos can be detected by Super-Kamiokande and IceCube which results in strong limits on spin-independent and spin-dependent interactions of dark matter \cite{Super-Kamiokande:2015xms, IceCube:2021xzo}\footnote{The limits from ANTARES \cite{ANTARES:2016xuh} are weaker than IceCube for the dark matter mass range considered in this paper}. In Ref. \cite{Bueno:2004dv}, the sensitivity of DUNE to DM annihilation to heavy quarks has been evaluated using electron neutrinos. As muons offer arguably better source pointing resolution, we estimate for the first time, the sensitivity of DUNE utilising the flux of muon neutrinos. We calculate the energy and angular cuts on the predicted atmospheric neutrino events that will favor the signal from dark matter annihilation in the Sun. For this, we have used the Monte Carlo program \texttt{NuWro} \cite{Golan:2012wx} to simulate the interactions of GeV-neutrinos with argon nuclei. We find that for the heavy-quark channel, the shoulder neutrinos provide better sensitivity than KDAR neutrinos in DUNE. However, the existing limits from upward-going muons in Super-Kamiokande are stronger than our projected sensitivity from contained events in DUNE. It would be interesting to study the sensitivity of DUNE from upward muons, but it is beyond the scope of this work. In \Cref{tab:bird}, we provide a bird's eye view of the detection capability of Super-Kamiokande, IceCube, and DUNE to the spike and shoulder neutrinos. We tabulate if the detector can reasonably reconstruct the incident neutrino energy and direction using spike or shoulder neutrinos. The spike neutrinos scatter via the quasi-elastic like (QEL) interactions, and the final state lepton is isotropic. As a result, the sensitivity of Super-Kamiokande and DUNE is obtained from the counting of total events in Ref. \cite{Rott:2015nma, Rott:2016mzs}. Recently, it was proposed that DUNE can reconstruct the neutrino direction using the final-state proton, which cannot be detected in Super-Kamiokande, and results in the reduction of atmospheric backgrounds \cite{DUNE:2021gbm}. The shoulder neutrinos mostly undergo deep-inelastic scattering (DIS) and the kinematics of the final-state lepton is correlated with the incident neutrino energy and direction. In Ref. \cite{Super-Kamiokande:2015xms}, Super-Kamiokande provides limits for elastic dark matter using both upward-muon and contained events from all-flavors. It is clear from Ref. \cite{Bell:2021esh} that the Super-Kamiokande sensitivity is dominated by the limits from upward muons. In our analysis, we map these limits on to the parameter space for inelastic dark matter. The constraints on inelastic dark matter above 100 GeV from IceCube is studied in Ref. \cite{Catena:2018vzc}

Due to the limitations of the Monte Carlo generator \texttt{NuWro}, we limit our analysis to dark matter mass below 100 GeV. In this range, there are competing limits from direct-detection experiments. We use the WimPyDD package \cite{Jeong:2021bpl} to estimate the relative event-rates for elastic and inelastic dark matter, and use the latest sensitivity of LUX-ZEPLIN (LZ) \cite{LUX-ZEPLIN:2022xrq}, XENONnT \cite{XENONCollaboration:2023orw}, PandaX-4T \cite{PandaX-4T:2021bab}, and the \C3F8 run of PICO-60 (PICO-\C3F8) \cite{PICO:2019vsc}. We find that these experiments provide a good coverage of the range of dark matter mass considered in this paper.    

In the next section, we present a brief overview of inelastic dark matter: direct detection and its gravitational capture in the Sun. The evaluation of the neutrino flux at Earth and at detectors from DM annihilation in the Sun in reviewed in \cref{sec:flux}. The atmospheric background and our approach to determining cuts to reduce the background relative to the signal of neutrinos from DM annihilation in the Sun appear in \cref{sec:atm}. In \Cref{sec:results} we present our results, while \cref{sec:conclusions} summarizes our main results and outlook. 

\section{Overview of Inelastic Dark Matter}\label{sec:idm}

In typical models of inelastic DM, the dark sector comprises two fermionic mass eigenstates $\chi_1$ and $\chi_2$.  Their masses differ by an endothermic mass splitting $\delta = M_2 - M_1 >0$. In these models, only off-diagonal interactions between the two mass states are allowed at the tree-level, i.e., $\chi_2$ couples to $\chi_1$ but the self-couplings $\chi_1 - \chi_1$ and $\chi_2 - \chi_2$ are forbidden. These self-couplings may be generated at loop-level \cite{Chauhan:2017eck}, and hence are sub-dominant. The cosmological dark matter density today is dominated by the lighter state $\chi_1$, however, there are circumstances when a sizable population of $\chi_2$ may also survive until today. In this work, we  do not make assumptions about the details of the particle physics model and only look at three parameters of inelastic DM --- the mass of lighter DM particle ($M_\chi$), mass splitting ($\delta$), and the DM-nucleon scattering cross-section ($\sigma_{\chi N}$). We will also assume isospin invariant couplings for simplicity. As a consequence of the mass splitting, a fraction of the energy budget of the DM interaction is spent to create the heavier state. This changes the kinematics of the interaction, and as a result, the available phase space for inelastic scattering is usually smaller than for elastic scattering. In this section, we quantify the impact of the inelasticity parameter $\delta$ on direct-detection rates as well as on gravitational capture of inelastic DM in Sun.

\subsection{Direct Detection}

Dark matter direct-detection experiments are designed to observe the energy transferred by galactic DM particles as they pass through the detector volume and interact with nuclei (or electrons) of the target material. Initially designed with a focus on the Weakly Interacting Massive Particle (WIMP) paradigm, most of these detectors are optimized to look for signals of a $\sim$100 GeV DM particle scattering via weak-scale interactions \cite{Goodman:1984dc}. The non-observation of any significant events in these experiments has significantly constrained the viable parameter space for WIMP-like DM \cite{PandaX-4T:2021bab, LUX-ZEPLIN:2022xrq, XENONCollaboration:2023orw}. Consequently, the scope of direct detection searches has broadened, especially at the low DM mass frontier \cite{SuperCDMS:2014cds, SuperCDMS:2015eex, SuperCDMS:2015lke, CRESST:2015txj, EDELWEISS:2016nzl, DarkSide:2018bpj,CRESST:2019jnq, PICO:2019vsc}. 

The rate of nuclear recoils (counts per kilogram per day) in the direct-detection experiments is given by, 
\begin{equation}\label{eq:Ndd}
    \mathcal{R}_{\rm DD} = \int_{E_R^{\rm min}}^{E_R^{\rm max}}dE_R\,\frac{\rho_\chi}{M_N M_\chi} \int_{v_{\rm min}}d^3v\,v\,f(\vec{v})\,\frac{d \sigma_{\chi N}}{dE_R}
\end{equation}
where $\rho_\chi \sim 0.3-0.4 \,\rm GeV/cm^3$ is the local dark matter density \cite{Catena:2009mf,Bovy:2012tw}, $M_N$ is the nucleon mass, $M_\chi$ is dark matter mass, $f(\vec{v})$ is the velocity distribution of dark matter in the vicinity of the Earth with $\int d^3v f(\vec{v})=1$, and $d \sigma_{\chi N}/dE_R$ is the differential DM-nucleon interaction cross section. The interaction cross section includes the form factors which are functions of the nuclear recoil energy $E_R$. The event rates in these detectors depends on the range of recoil energies to which the detector is sensitive and on the minimum dark matter velocity required to generate detectable nuclear recoils ($v_{min}$). 

As none of the direct-detection experiments have observed a significant excess in nuclear recoils above their background expectations, \Cref{eq:Ndd} can be used to determine their exclusion limits on elastic DM parameters $M_\chi$ and $\sigma_{\chi N}$. The sensitivity of these experiments depends on the velocity distribution of DM in the vicinity of the Earth. It is usually assumed that the DM has a Maxwellian velocity distribution in the galactic frame with $v_0$=220 km/s and cutoff at 550 km/s, which is the escape velocity of our galaxy. As the Earth is moving through the DM halo, $f(v)$ in \Cref{eq:Ndd} is the time-averaged distribution in the \emph{boosted} frame. 

In the case of inelastic dark matter, the minimum DM velocity required to produce a specific recoil on a nucleus of mass $M_N$ is increased due to the the mass splitting, 
\begin{equation}
	v_{\rm min} (E_R) = \frac{1}{\sqrt{2 M_N E_R}} \left( \frac{M_N}{\mu_{\chi N}}E_R + \delta \right). 
\end{equation}
where $\mu_{\chi N}=M_{\chi} M_N/(M_\chi+M_N)$ is the reduced mass of dark matter nucleus system. As a result, the available phase space for inelastic dark matter is typically smaller when compared to the elastic scattering $(\delta=0)$, and the event rates are relatively smaller. Not only that, inelastic DM scattering is also more sensitive to the high-velocity tail of the distribution. 

\begin{figure}[t]
    \centering
    \includegraphics[width=0.49\textwidth]{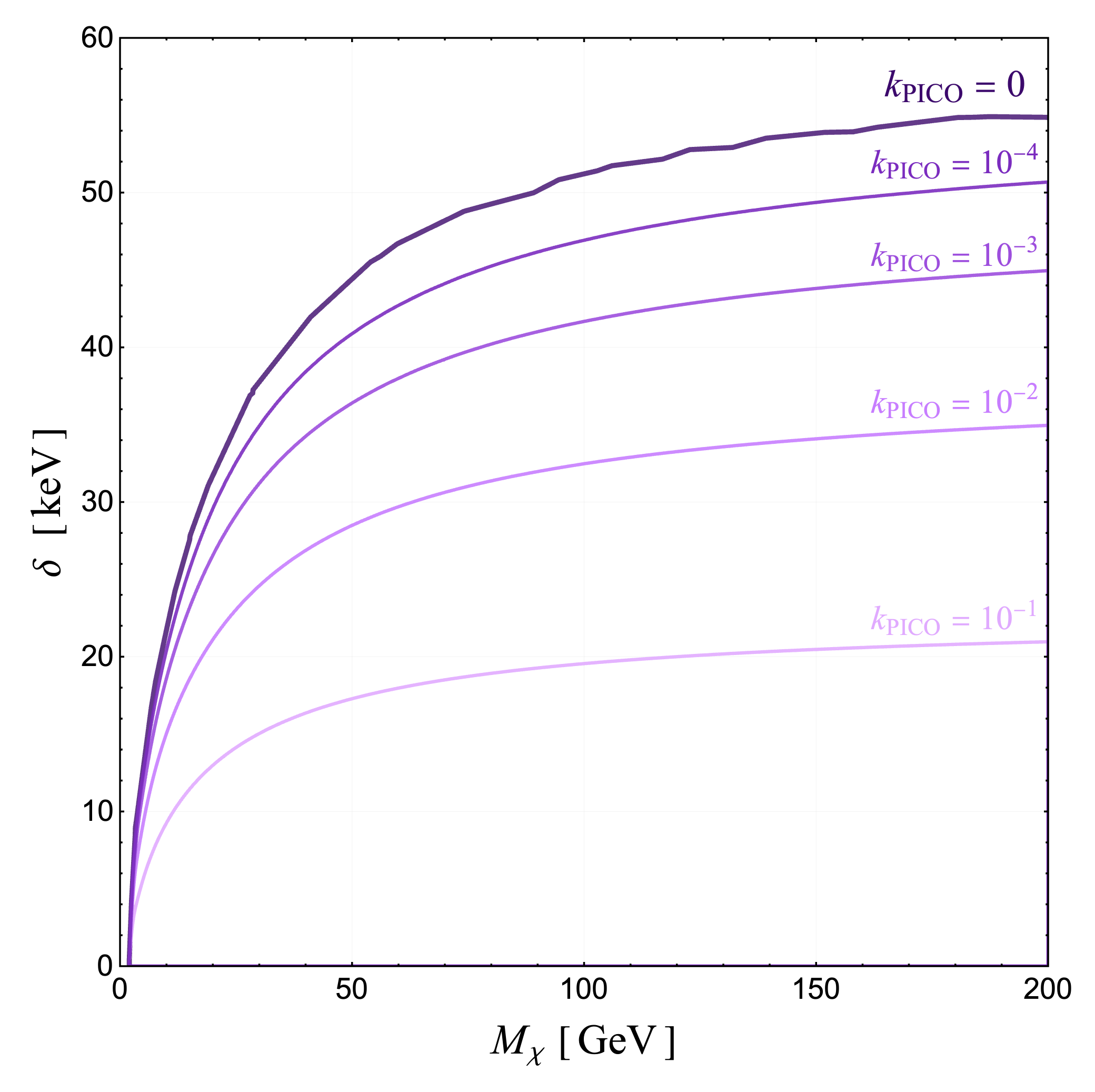}
    \includegraphics[width=0.49\textwidth]{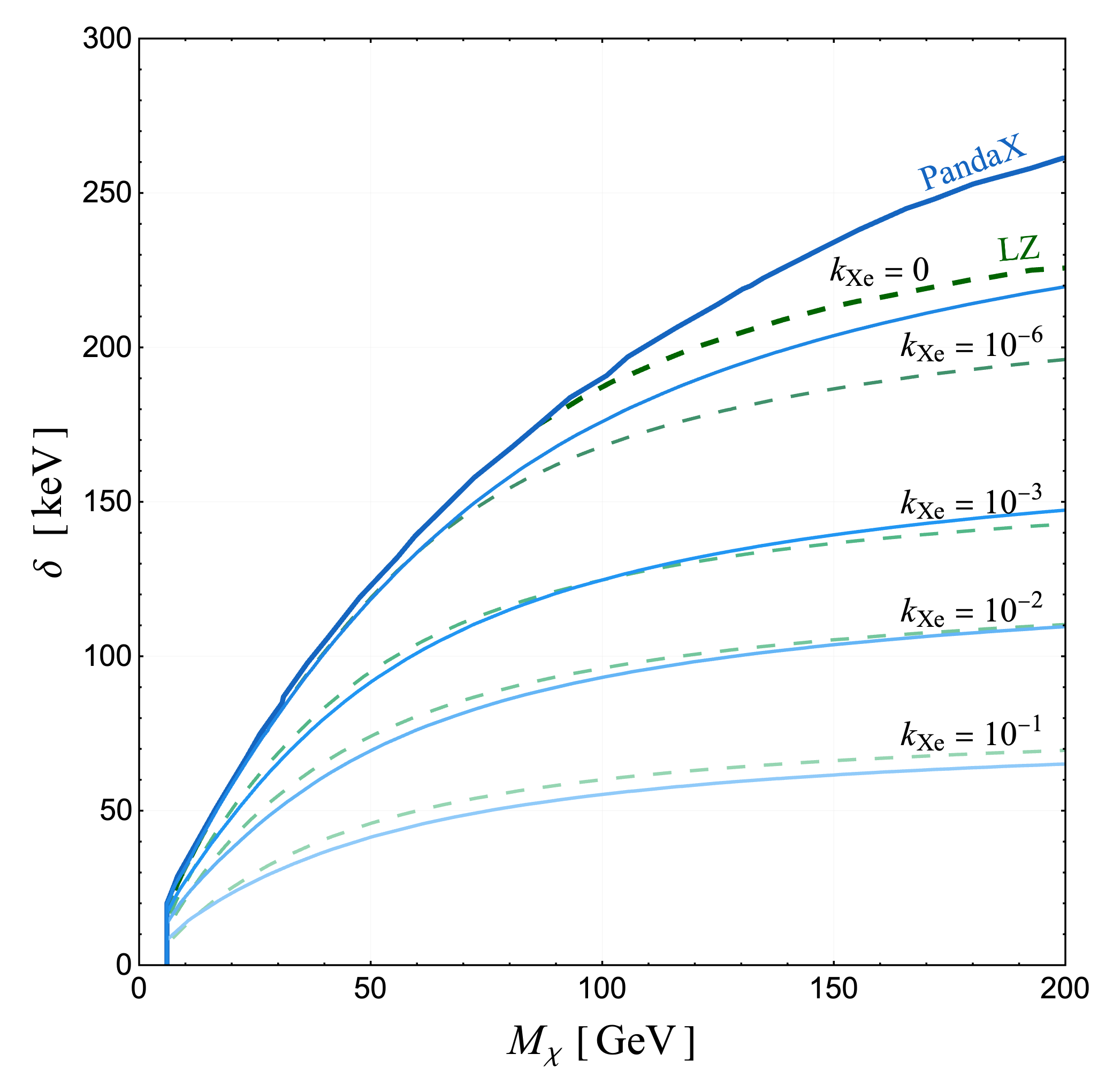}
    \caption{\label{fig:dd} The contours for relative event rates $k_{\rm DD}$ (see \Cref{eq:kdd}) are shown for PICO-\C3F8 (left) and the Xenon-based experiments (right). Note the difference in the limits of the ordinate between figures. The contours for LZ (dashed) and PandaX (solid) are different as PandaX is sensitive to larger nuclear recoils. The difference is significant for DM mass above $\sim$75 GeV and mass splitting above $\sim$150 keV.}
\end{figure}

We use WimPyDD \cite{Jeong:2021bpl} to calculate the event rates for various direct-detection experiments supported by the package. WimPyDD can evaluate the event rates for elastic as well as inelastic dark matter for the Xenon-based experiments (XENONnT, and LZ) and the bubble chamber detector PICO-60 with the \C3F8 target. We have also evaluated the sensitivity of PandaX which differs from XENON and LZ primarily due to increased sensitivity at large recoils \cite{PandaX-4T:2021bab}. Although other target nuclei are supported, a complete implementation of other detectors such as CRESST \cite{CRESST:2019jnq}, DarkSide \cite{DarkSide:2018bpj}, and CDMS \cite{SuperCDMS:2015lke} in WimPyDD is beyond the scope of this work. As the sensitivity obtained by WimPyDD is only approximate, it is advantageous to utilize the relative event-rate 
\begin{equation}\label{eq:kdd}
	k_{ \rm DD}(M_\chi, \delta) = \mathcal{R}_{ \rm DD}(M_\chi, \delta)\,/\,\mathcal{R}_{\rm DD}(M_\chi, 0)	\,,
\end{equation}
which is less sensitive to inaccuracies of the package, and represents the scaling of the event rate with the mass-splitting parameter $\delta$. Using $k_{ \rm DD}$, we obtain the limits on the inelastic DM-nucleon cross-section from a direct-detection experiment by scaling the limits for elastic DM scattering. The exclusion limits for inelastic DM is thus given by, 
\begin{equation}
    \sigma_{\rm DD}^{\rm lim}(M_\chi, \delta) = \sigma_{\rm DD}^{\rm lim}(M_\chi, 0) \times  k_{\rm DD}(M_\chi, \delta)\, ,
\end{equation}
where $\sigma_{\rm DD}^{\rm lim}(M_\chi, 0)$ is the current limit on the spin-independent elastic scattering cross-section. In future, one can use $k_{\rm DD}$ to obtain new limits when there is an update from the experiment. In \Cref{fig:dd}, we show the contours in $M_\chi-\delta$ parameter space for fixed $k_{\rm DD}(M_\chi,\delta)$ for PICO-\C3F8 and the Xenon-based experiments LZ and PandaX. We find that for a given DM mass $M_\chi$, there exists $\delta_{\rm max}$ beyond which a direct-detection experiment is completely insensitive to DM-nucleon scattering. Moreover, $\delta_{\rm max}$ also depends on the maximum recoil threshold as evident in the larger sensitivity of PandaX when compared to LZ and XENON. Similar conclusions have been obtained in Ref. \cite{Song:2021yar}. At the time of this writing, XENON and PandaX provide the leading limits for DM mass between 5--9 GeV whereas LZ is most sensitive above 9 GeV. For DM mass between 2--5 GeV, we use the limits by PICO-\C3F8 \footnote{In this range, the liquid argon based experiment, DarkSide \cite{DarkSide:2018bpj}, provides leading sensitivity today and evaluating the relative event-rate on argon is interesting. However, DM with mass less than 4 GeV is evaporated from the Sun and the marginal improvement does not warrant a dedicated implementation in WimPyDD}. Recently, the PICO collaboration presented their limits on inelastic dark matter in \cite{PICO:2023uff}, and our results obtained with \Cref{eq:kdd} are in agreement.

It is worth mentioning that other experiments like CRESST have enhanced sensitivity for larger mass splittings, and future heavy-nuclei based experiments can also probe a wider DM parameter space \cite{Song:2021yar}. Moreover, the sensitivity of these experiments is also improved by accounting for an extra-galactic component in the DM velocity distribution \cite{Herrera:2023fpq}. However, systematic inclusion of this high-velocity component in WimPyDD as well as gravitational capture in the Sun (as discussed below) is beyond the scope of this work and left as a future effort. 

\subsection{Gravitational Capture in Sun}

\begin{figure}[t]
    \centering
    \includegraphics[width=0.48\textwidth]{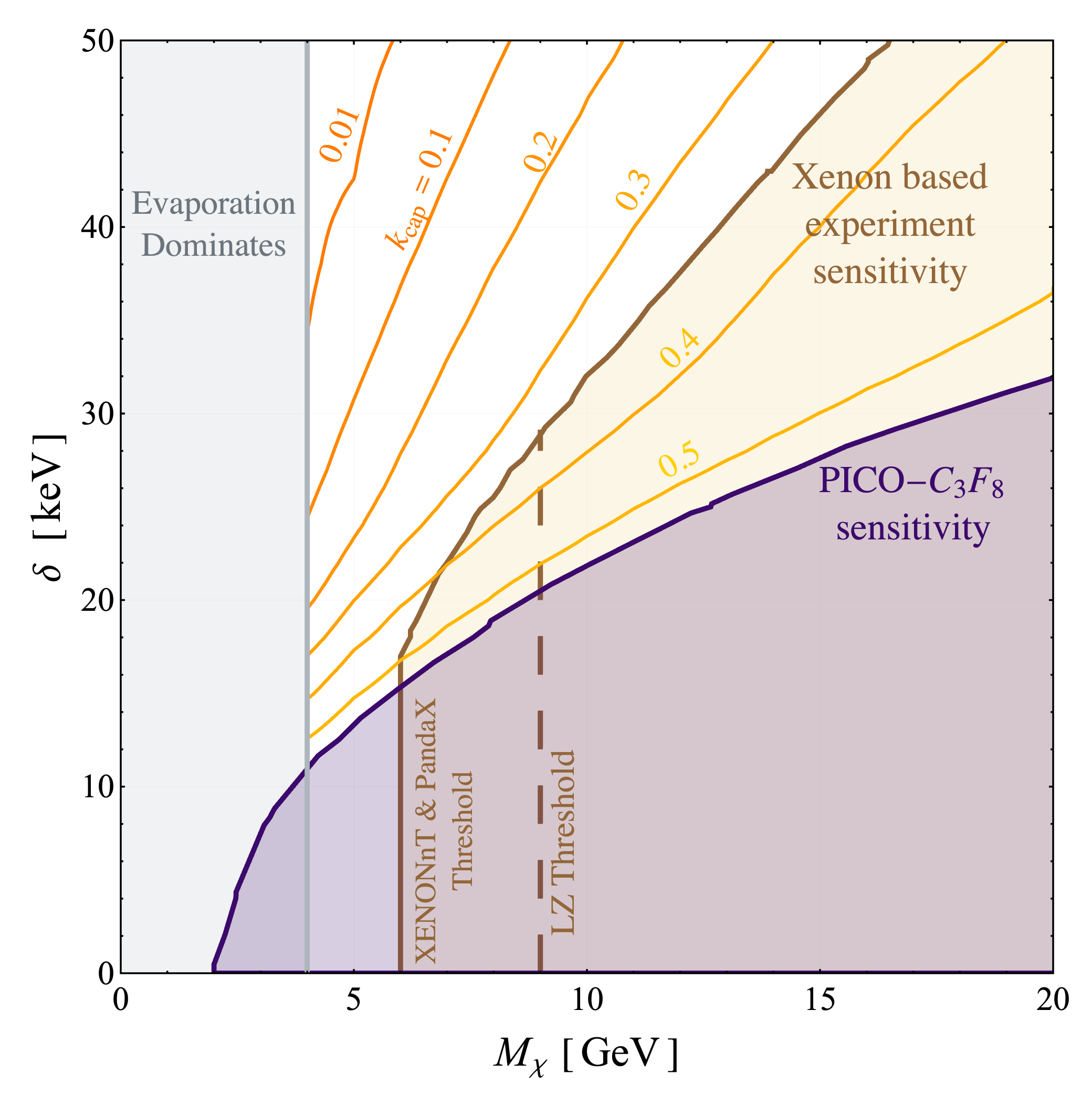}
    \includegraphics[width=0.49\textwidth]{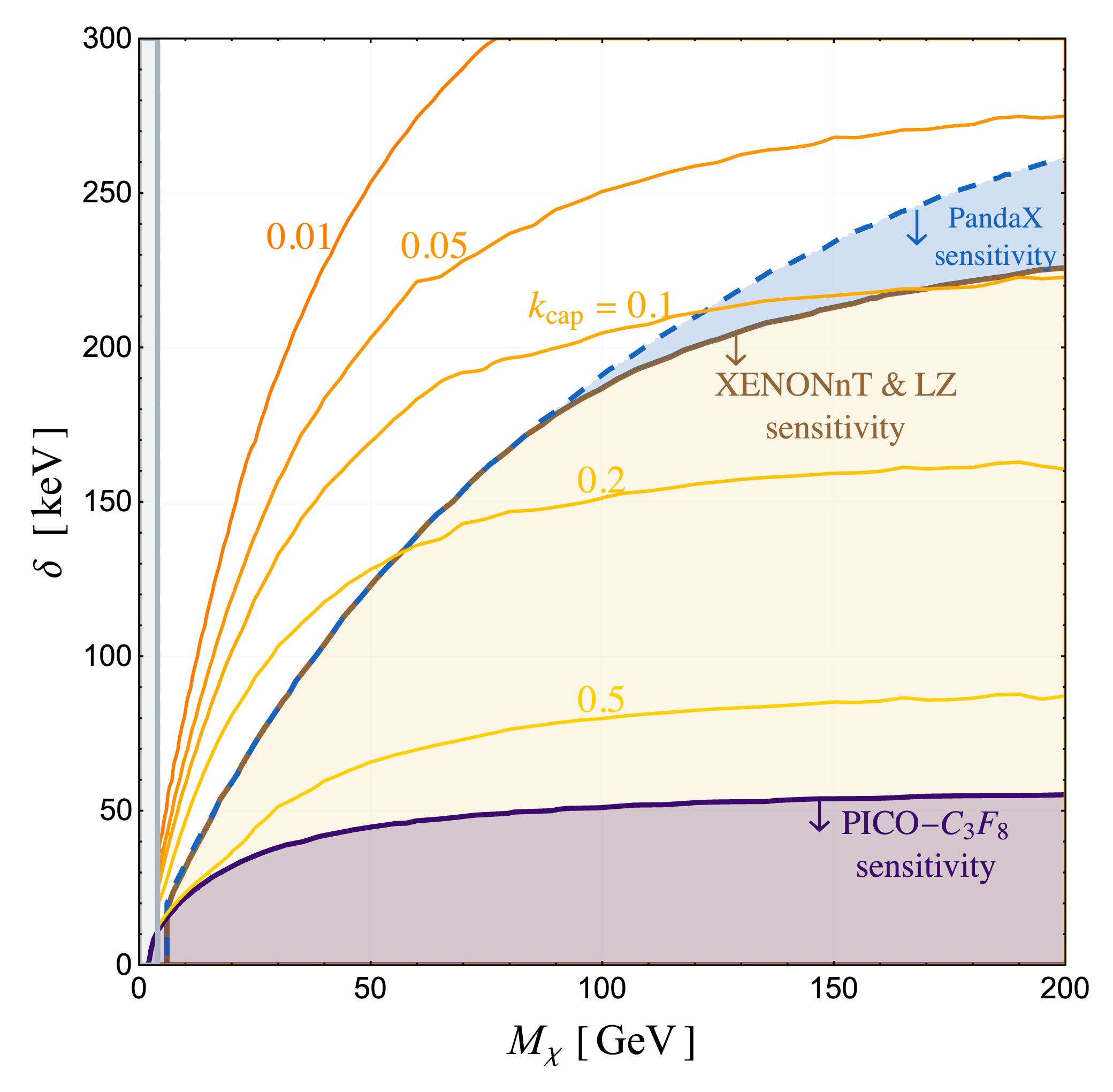}
    \caption{\label{fig:cap} The contours for relative capture rate $k_{\rm cap}$ (see \Cref{eq:kcap}) are compared with the sensitivity of direct-detection experiments. The figure on the left focuses on the small-mass parameter space. The shaded regions show non-zero relative event rate in direct-detection experiments, $k_{\rm DD} \geq 0$, for PICO-C3F8 (purple), XENON and LZ (cream), and PandaX (blue). The gray shaded region represents the evaporation mass of dark matter, which we fix at 4 GeV for all $\delta \geq 0$. One can see that there is a significant parameter space where direct-detection experiments are not sensitive, but where indirect detection from a significant gravitational capture and annihilation of DM in the Sun is possible.}
\end{figure}

When a DM particle from the galactic halo falls into the gravitational potential of the Sun, it accelerates to a velocity $w(r) = \sqrt{u^2 + v_{esc}^2(r)}$, where $u$ is the DM velocity outside the gravitational potential and $v_{esc}$ is the local escape velocity, both in units of $c$ (natural units). The inelastic interaction between incident DM and a target nucleus in the Sun is kinematically allowed only if $w^2(r) > 2 \delta / \mu$ where $\mu$ is the reduced mass of the DM--nucleus system \cite{Nussinov:2009ft, Menon:2009qj}. As a result, for typical values of the mass splitting, only the high-velocity tail of DM distribution results in a gravitational capture. The differential capture rate by the $i$-th isotope in the Sun is given by,
\begin{equation}
    \label{eq:cap}
    \frac{dC_i}{dV} = n_\chi n_i(r) \int du \frac{f(u)}{u}w(r)\int dE_R \frac{d\sigma}{dE_R}\,,
\end{equation}
where $f(u)$ is the DM velocity distribution boosted to the frame of the Sun, $n_\chi = \rho_\chi / M_\chi$, and $n_i(r)$ is the local number density of the target isotope. The limits for integration over the recoil energy and the velocity are non-trivial as the phase-space for capture is significantly reduced due to the mass splitting $\delta$. We follow the method outlined in Ref. \cite{Blennow:2015hzp} for evaluating the integrals in \Cref{eq:cap}. The \emph{total} rate of capture of DM in Sun is given by, 
\begin{equation}
    \Gamma_{\rm cap} = \sum_{i} \int_{0}^{R_{\rm sun}}4 \pi r^2 dr \frac{dC_i}{dV}\,,
\end{equation}
where we sum over all 29 isotopes given by the B16-AGSS09 solar model \cite{Vinyoles:2016djt}. As the capture rate for inelastic dark matter with spin-dependent interactions is very small \cite{Menon:2009qj, Blennow:2015hzp}, we only look at spin-independent scatterings of dark matter. Moreover, we assume that the number density of captured dark matter in the Sun has reached equilibrium and the rate of annihilation is given by the steady state solution, 
\begin{equation}
    \Gamma_{\rm A} = \frac{\Gamma_{\rm cap}}{2}. 
\end{equation}
Lastly, the evaporation of low-mass dark matter will be affected by a non-zero mass splitting \cite{Blennow:2018xwu}. We assume that any change in reduced evaporation would be dominated by the reduction in capture rate and do not consider dark matter mass below 4 GeV \cite{Gould:1987ju, Garani:2021feo}. In detailed models, a small elastic-scattering cross section is always generated at one-loop level which aids in attaining equilibrium and contributes to evaporation. 

Similar to the application of direct-detection limits on elastic DM to inelastic DM, we are interested in the relative reduction in the capture rate due to the inelastic DM mass splitting, so we determine
\begin{equation}\label{eq:kcap}
	k_{\rm cap} =  \rm \Gamma_{cap}(M_\chi, \delta)\,/\,\Gamma_{cap}(M_\chi, 0)\,.
\end{equation}
In \Cref{fig:cap}, we show the contours of constant $k_{\rm cap}$ in the $M_\chi - \delta$ parameter space. As expected, the inclusion of the mass splitting yields a smaller capture rate as compared to the elastic DM interactions (i.e., $\delta=0$ line in the figure). We also show the regions of parameter space where complementary limits from direct-detection experiments exist from Xenon-based experiments as well as from PICO-\C3F8. It is interesting to note that there exists a region of parameter space that is inaccessible to direct-detection experiments but does not suffer from an exponential reduction in capture rate. In this region, even though the capture rate is reduced by factor of $0.1-0.01$, there are no competing limits from direct-detection experiments. The only way to probe this space is through neutrino detection experiments. In Ref. \cite{Kang:2023gef} discusses the complementarity between direct-detection experiments and capture in the Sun of inelastic dark matter to obtain halo-independent bounds. 
\section{Neutrino Flux from DM annihilation in Sun}
\label{sec:flux}

\begin{figure}
    \centering
    \includegraphics[width=0.49\linewidth]{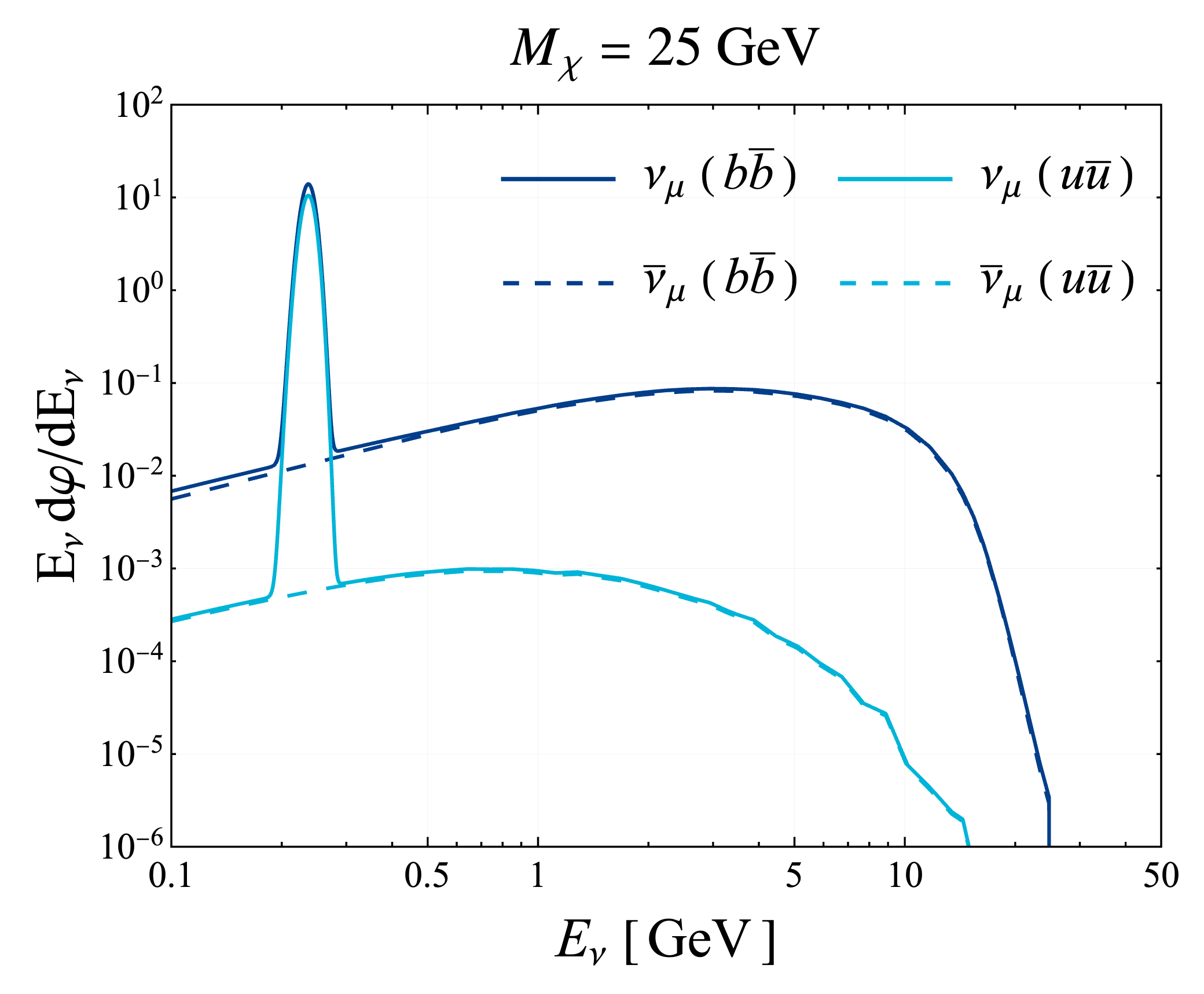}
    \includegraphics[width=0.49\linewidth]{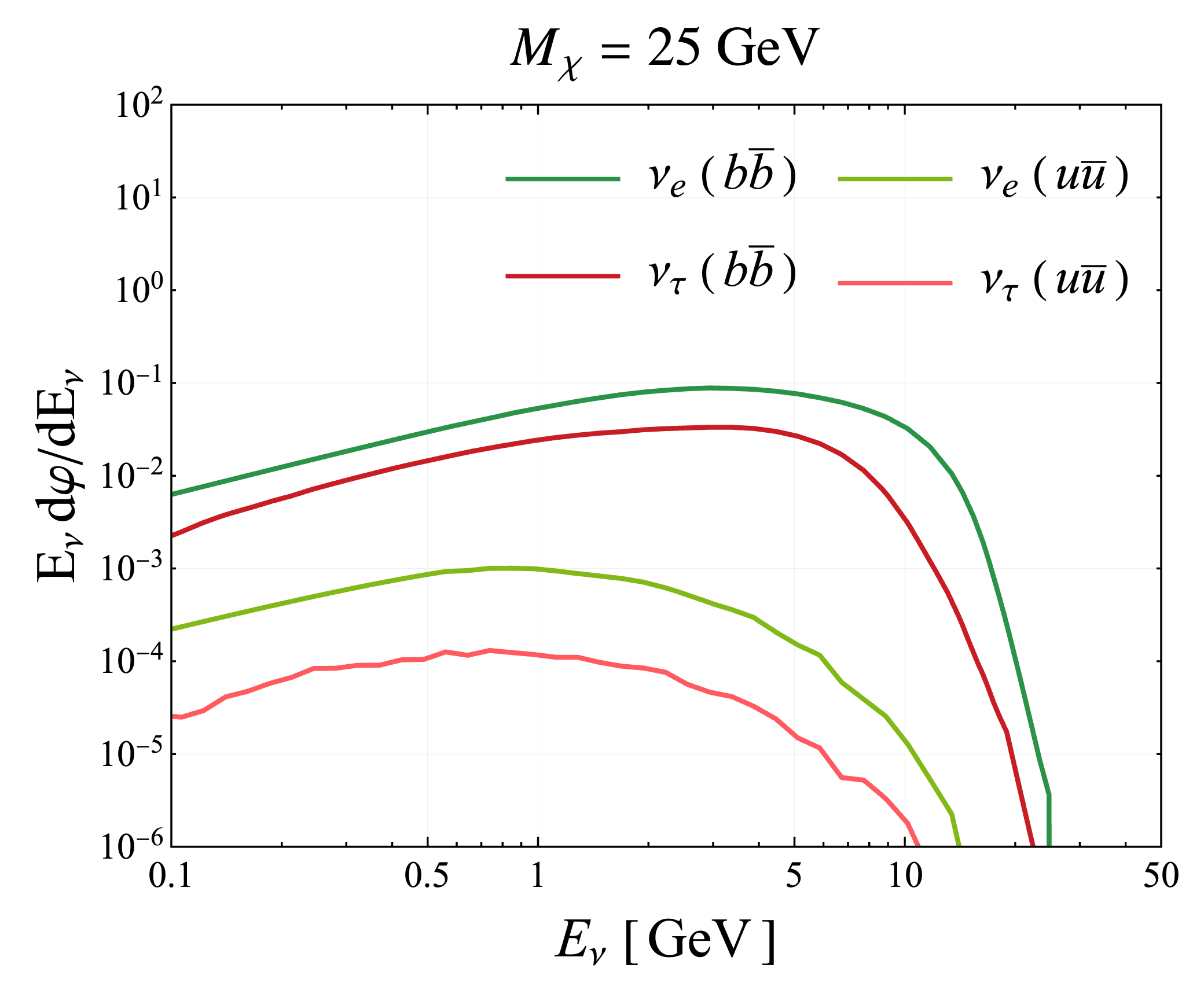}
    \caption{\emph{Left:} The energy-scaled spectrum of muon neutrinos and anti-neutrinos per DM annihilation ($ E_\nu\,d \varphi_\beta/dE_\nu$) in the Sun is shown for a benchmark DM mass of 25 GeV and for two annihilation channels - `$b\bar{b}$' (blue) and `$u\bar{u}$' (light blue). The neutrino (anti-neutrino) spectrum is shown with solid (dashed) curve. The \emph{spike} in the $\nu_\mu$ spectrum is due to primary and secondary $K^+$ that decay-at-rest (KDAR) in the Sun. For illustration, we represent the mono-energetic feature with a Gaussian of width 10 MeV which is the typical energy resolution of a LArTPC detector at these energies. \emph{Right:} The energy-scaled spectrum of non-muon flavors is shown. The spectrum of anti-neutrinos is same as neutrinos. \emph{Both:} The broad feature from $E_\nu=500$ MeV to $E_\nu=M_\chi$ in all neutrino spectra, called the \emph{shoulder}, is significant only for the heavy-quark channel. There are additional features in the neutrino spectra at below 250 MeV, notably from three-body decays of kaons, which are not shown here.}
    \label{fig:sample}
\end{figure}

In a simplified and minimal model of inelastic DM, it is reasonable to assume that the gravitationally captured DM in the Sun annihilates to standard model particles. In hadrophilic models, the DM annihilates to a quark-antiquark pair which hadronize in the solar media. The subsequent decays of the mesons result in neutrinos that escape the Sun. In detailed models of DM, it is also possible that the dominant annihilation channel is to leptons or beyond standard model particles \cite{Smolinsky:2017fvb}, however, we do not consider these models here. We also ignore the details of the mediator and only assume scalar contact interactions.  

The flux of neutrinos of flavor $\alpha$ at a detector on Earth is given by, 
\begin{equation}\label{eq:flux}
	\frac{d \Phi_\alpha^{\rm D}}{dE_\nu} = \frac{\Gamma_{\rm A}}{4 \pi d^2} \sum_{\beta} P_{\alpha \beta}  \, \frac{d \varphi_\beta}{dE_\nu}
\end{equation}
where, $d \varphi_\beta/dE_\nu$ is the energy-spectrum of neutrinos produced per annihilation and $P_{\alpha \beta}$ is the average conversion probability which includes propagation inside the Sun, through vacuum to the Earth, and propagation inside the Earth to the detector location. In this work, we assume that $d=1$ A.U. is the average distance between the Sun and Earth, and we ignore the effects of the  eccentricity of the Earth's orbit. We use the publicly available package PPPC\,4\,DM$\nu$ \cite{Baratella:2013fya} to estimate the neutrino flux as well as the DM annihilation rate, $\Gamma_{\rm A}$. The spectrum of neutrinos of various flavors at the source, without any flavor conversion effects, is shown in Figure \ref{fig:sample} for a benchmark DM mass of 25 GeV for two different annihilation channels, to `$u\bar{u}$' and  to `$b\bar{b}$'. Note that there are two distinct features: a mono-energetic \emph{spike} in the $\nu_\mu$ spectrum, and a broad \emph{shoulder} in all six components ($\nu_e$, $\bar{\nu}_e$, $\nu_\mu$, $\bar{\nu}_\mu$, $\nu_\tau$, and $\bar{\nu}_\tau$). The spike at 236 MeV arises from primary and secondary $K^+$ that decay at rest \cite{Rott:2012qb, Bernal:2012qh}, whereas the shoulder is mostly due to prompt decays of primary mesons produced after hadronization. For the spike, the integrated form of \Cref{eq:flux} can be obtained as, 
\begin{equation}\label{eq:spike_lim}
    \Phi^{\rm spike} = \frac{\Gamma_{\rm A}}{4 \pi d^2} \times P_{\mu \mu} \times \frac{2 M_\chi r_K}{2 m_K} \mathcal{B}
\end{equation}
where $r_K$ is the energy fraction that yields kaons at rest \cite{Rott:2015nma}, $m_K \sim 494$ MeV is the mass of charged kaon, and $\mathcal{B}=0.68$ is the charged kaon branching fraction to mono-energetic $\nu_\mu$. In Ref. \cite{DUNE:2021gbm}, the DUNE collaboration reported the projected sensitivity of the flux of these 236 MeV neutrinos in a DM model independent form as, 
\begin{equation}\label{eq:spike_lim_dune}
    \Phi^{\rm spike} \, (400 \, \text{kton-yr}) \leq 10^2 \, \rm m^{-2}\,s^{-1}
\end{equation}
translates to limits on the DM capture rate in the Sun. For an honest comparison, we use the sensitivity from generator-level analysis without accounting for the detector effects in reconstruction, which reduces the sensitivity by an order or magnitude. The sensitivity of DUNE to shoulder neutrinos from DM annihilation is the Sun is the main subject of this paper.  

One should note that the shoulder is significant only if DM annihilates to heavy quarks, whereas the spike is present in both light-quark and heavy-quark channels \cite{Rott:2015nma}. As we do not discuss the detailed model of inelastic DM, we consider these two channels, $u\bar{u}$ and $b\bar{b}$, independently as benchmark annihilation channels with unit branching fraction. Moreover, we do not consider channels such as $W^+W^-$ and $t \bar{t}$ as the main focus of the paper is on DM masses below 100 GeV. Lastly, we do not explicitly consider leptonic channels such as $\tau \bar{\tau}$ and $\nu \bar{\nu}$ as the results for these channels can be analogously obtained. 

\begin{figure}[b]
	\centering
	\includegraphics[width = 0.49 \linewidth]{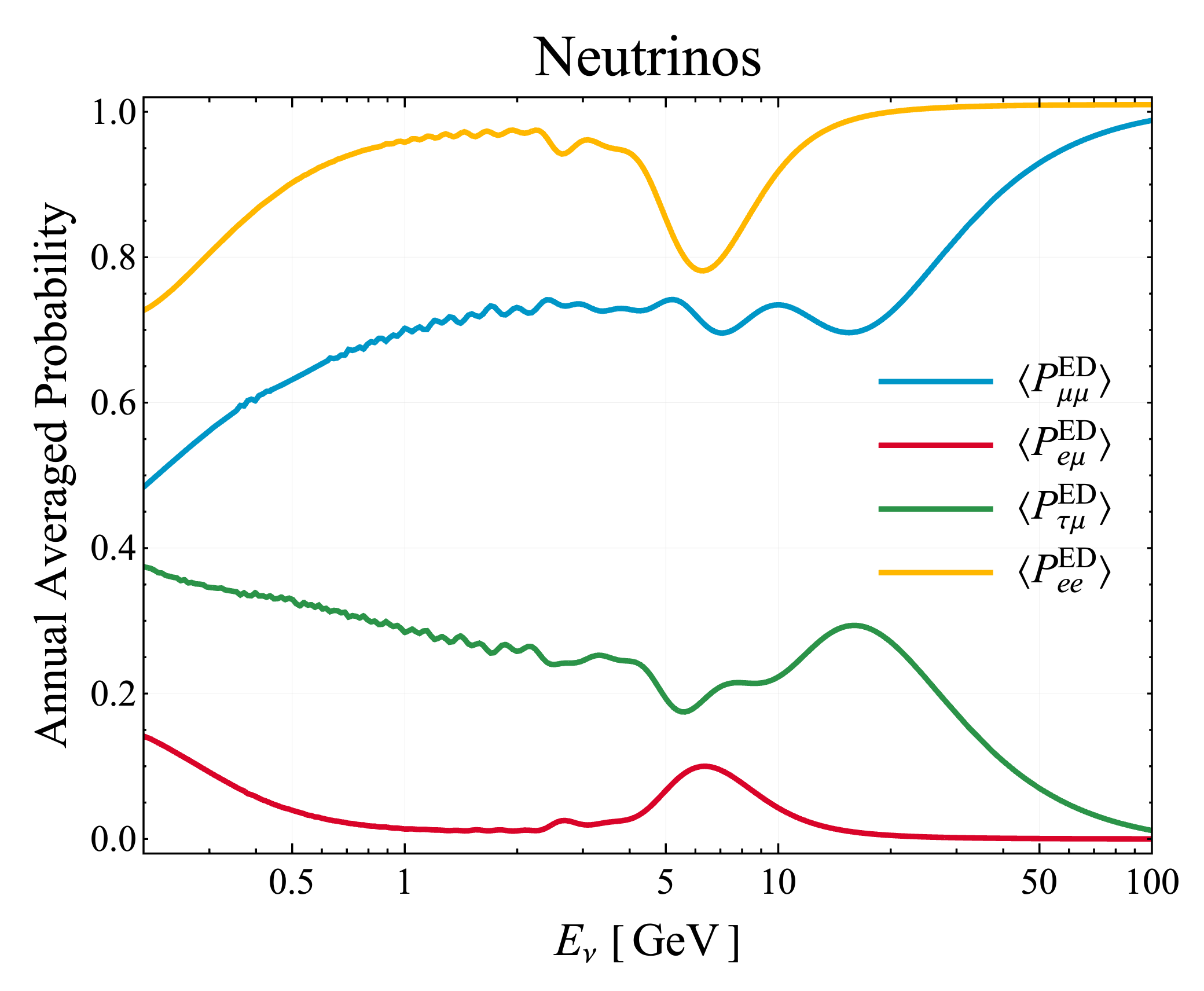}
	\includegraphics[width = 0.49 \linewidth]{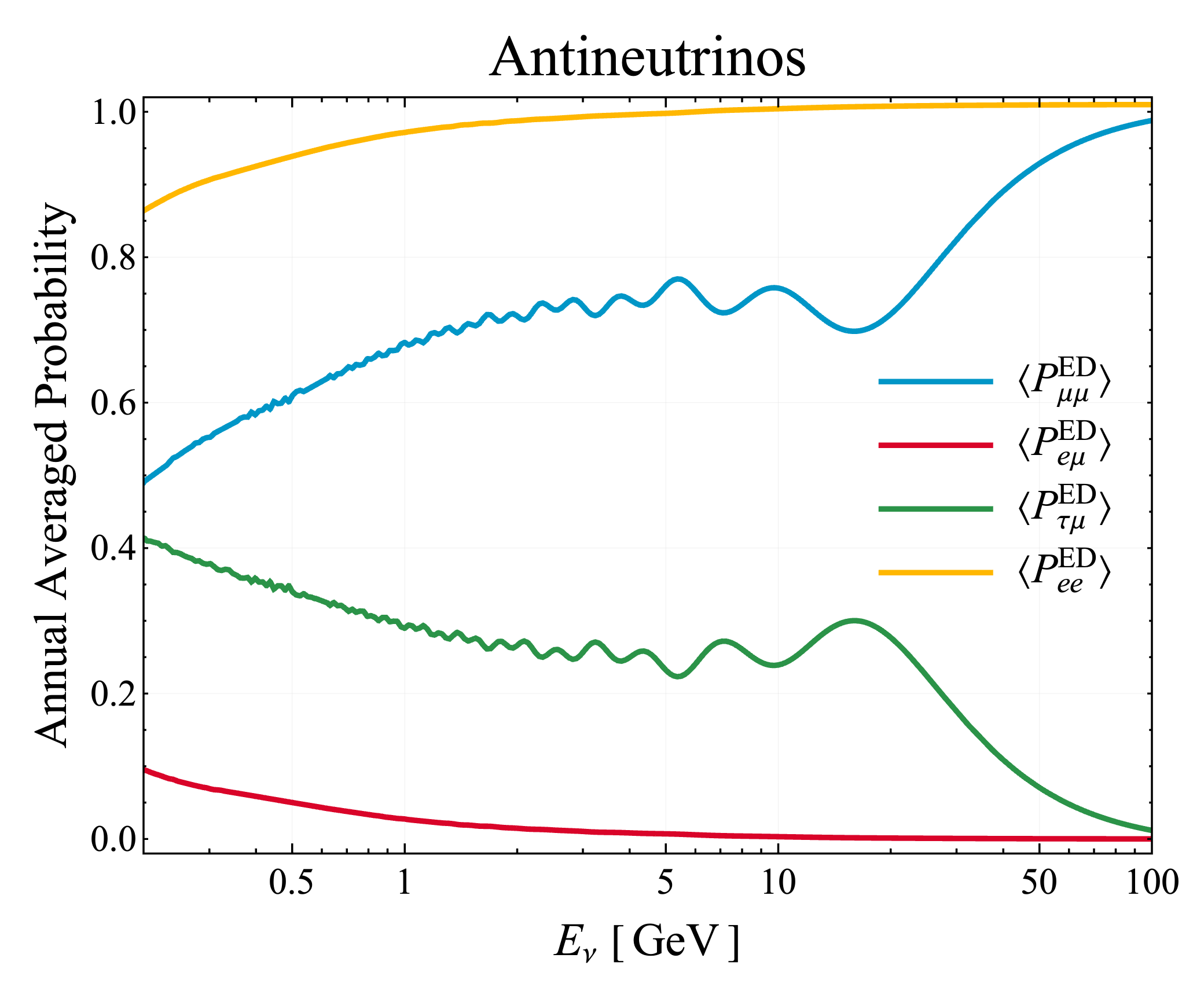}
	\caption{\label{fig:ann_avg} The flavor conversion probabilities, averaged over the position of the Sun over a period of one year at the proposed location of DUNE far-detector,  are shown for neutrinos (left) and antineutrinos (right) as functions of energy.}
\end{figure}

We quantify the effect of flavor conversion in the Earth using the latest neutrino oscillation parameters from NuFIT\cite{Esteban:2020cvm}. We use the nuCraft software \cite{Wallraff:2014qka} to evaluate the flavor conversion probabilities due to vacuum-mixing as well as Earth-matter effects \cite{Wolfenstein:1977ue, Mikheyev:1985zog}. The parametric resonance \cite{Akhmedov:1988kd,Krastev:1989ix}, which is important for Earth-crossing neutrinos in the energy range 3--10 GeV \cite{Kelly:2021jfs}, is included in nuCraft. We evaluate the neutrino flux at 1 A.U. which includes flavor conversion in the Sun and in vacuum using PPPC\,4\,DM$\nu$. This flux without Earth-crossing, $d\Phi_\mu^{\rm E}/dE_\nu$, is multiplied by flavor conversion probabilities depending on the zenith angle of the Sun at a given time and location on Earth. As we are interested in event-rates in detectors over a span of several years of runtime, we average over the position of the Sun over the period of one year and obtain an annual averaged flavor conversion probability. Subsequently, the (annually-averaged) flux of $\nu_\alpha$ at a detector location is obtained by,  
\begin{equation}
    \frac{d\Phi_\alpha^{\rm D}}{dE_\nu} = \frac{d\Phi_e^{\rm E}}{dE_\nu} \times \langle P^{ED}_{e\alpha}\rangle + \frac{d\Phi_\mu^{\rm E}}{dE_\nu} \times \langle P^{ED}_{\mu\alpha}\rangle + \frac{d\Phi_\tau^{\rm E}}{dE_\nu}\times \langle P^{ED}_{\tau\alpha}\rangle 
\end{equation}
where $\langle P_{\alpha\beta}\rangle$ represents the conversion probabilities averaged over the zenith angles of the Sun. The antineutrino flux at the detector can be obtained analogously. We use AstroPy \cite{Astropy:2013muo} to calculate the zenith-angles of the Sun ($\psi$) at the proposed location of the DUNE far-detector, and we calculate $P_{\alpha \beta}^{ED}(E_\nu, \psi)$ using nuCraft for $100\times365$ samples of $\psi$ over a period of one year, before averaging. Our primary interests are the $\nu_\mu$ and $\bar\nu_\mu$ fluxes from the shoulder since the final-state charged leptons in the charged-current interactions are best candidates to identify the direction of the source, and $\mu^\pm$ offer the best angular resolution in DUNE. We evaluate $\langle P_{\alpha \beta}\rangle$ for neutrinos and antineutrinos and present  them in Figure \ref{fig:ann_avg}.

\section{Atmospheric Neutrino Background}
\label{sec:atm}

\begin{figure}[b]
	\centering
        \includegraphics[width=0.5\linewidth]{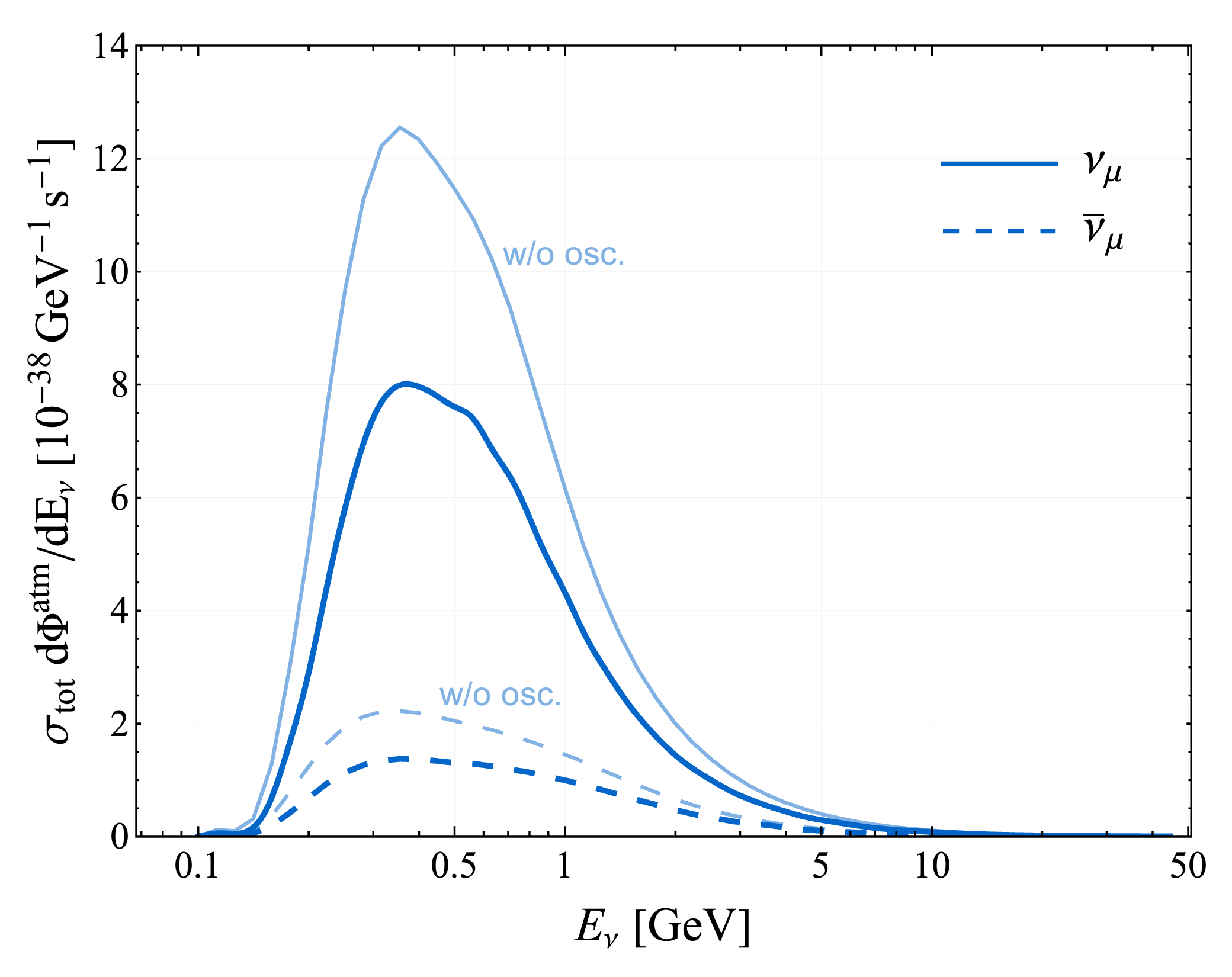}
	\caption{\label{fig:atm_int_nu} We show the product of the angle-integrated fluxes of atmospheric $\nu_\mu$ (solid) and $\bar{\nu}_\mu$ (dashed) and charged-current cross sections at DUNE. We have explicitly shown the effect of flavor-conversion in Earth: the lighter blue curves show the same $\nu_\mu$ and $\bar\nu_\mu$ fluxes times cross sections without neutrino oscillations in the Earth. Even though the fluxes are similar, $\bar{\nu}_\mu$ contribute less to the event rate due to relatively smaller cross section.}
\end{figure}

As discussed in the previous section, our focus is on low-mass DM particles which produce neutrinos with energies ranging from 500 MeV to 100 GeV. At these energy scales, one of the key challenges is to efficiently identify and reject the significant background arising from atmospheric neutrinos. Atmospheric neutrinos, originating from interactions between cosmic rays and the Earth's atmosphere, reach the detector from all directions \cite{Honda:2015fha}. A fraction of them can appear to come from the direction of the Sun. For a typical detector exposure that we consider, these neutrinos cannot be distinguished from the neutrinos originating from DM annihilation in the Sun, and thus limit the sensitivity of any terrestrial detector. In this section, we take a close look at the expected distributions of atmospheric neutrino events in DUNE and propose event selection criteria that would significantly reduce the atmospheric neutrino background.

As detailed estimates of the atmospheric neutrino flux at the location of the DUNE far-detector are unavailable, we adopt the flux predictions for Super-Kamiokande provided by Honda et al.\,\cite{Honda:2015fha}. This is a reasonable assumption as the proposed site for the DUNE is at a similar latitude to Super-Kamiokande \cite{Kelly:2021jfs}. The effect of atmospheric neutrino flavor conversion due to propagation in the Earth is evaluated using nuCraft\,\cite{Wallraff:2014qka} using the best-fit values of neutrino oscillation parameters and assuming the Normal Hierarchy of neutrino masses. As noted above, to obtain the energy-dependent cross sections as well as differential distributions, we use the neutrino Monte Carlo generator \texttt{NuWro} that simulates neutrino and antineutrino interactions with argon nuclei. Although Ref. \cite{Honda:2015fha} provides zenith and azimuth dependent flux, we use the all-sky-averaged flux as input to \texttt{NuWro}, and we assume that the distribution is isotropic at the energy scales in consideration. Similar to Ref. \cite{Rott:2016mzs}, we use the spectral function approach to model the nuclei and set \texttt{nucleus\textunderscore target}=2 and \texttt{sf\textunderscore method}=1 in the \texttt{NuWro} parameter initialization file. Other parameters are set to their default values. We only look at charged-current interactions. All neutral-current interactions channels are switched off. ITo obtain generalized results that are utilized later for the neutrino flux from DM annihilation in Sun, we also separately simulate interactions from mono-energetic $\nu_\mu$ and $\bar{\nu}_\mu$ with energies between 100 MeV and 100 GeV. As mentioned earlier, we only focus on the muon neutrinos and antineutrinos as we expect better source-pointing resolution for tracks as opposed to cascade-like events from electron and tau flavors.

Our focus is directed towards interactions that result in muon-like events, where the interaction point is situated within the volume of the detector. This is different from Super-Kamiokande, which primarily utilizes upward-going muons that mostly originate in the rock outside the detector \cite{Super-Kamiokande:2011wjy}. The upward muons are advantageous as they have much larger target volume, whereas the \emph{contained} events in DUNE carry more information about the incident neutrinos. Due to low thresholds and excellent particle identification capabilities of LArTPCs, the events will have additional detectable final-state particles (like protons and pions) that can help with reconstruction, however, we do not utilise them. The differential event rate for contained events from charged-current interactions of atmospheric muon neutrinos can be expressed as, 
\begin{equation}
	\frac{dN^{\rm atm}}{dE_\nu} =  \frac{M_{\rm target }\, T}{\text{kton}} \frac{N_A \times 10^9}{A} \times \sigma_{\rm tot}\frac{d\Phi^{\rm atm}}{dE_\nu}
\end{equation}
where $\sigma_{\rm tot}$ is the total charged-current cross section from all channels, $M_{\rm target }$ is the fiducial mass of the target nuclei (atomic mass $A$), and $T$ is the exposure time. In \Cref{fig:atm_int_nu}, we show the flux of atmospheric $\nu_\mu$ and $\bar{\nu}_\mu$ folded with their energy-dependent interaction cross section, with and without accounting for flavor conversions in the Earth. As DUNE is not expected to be magnetized, it cannot differentiate between $\mu^-$ and $\mu^+$, and the detector is only sensitive to the aggregate flux. We expect around 2300 events per 34 kton-yr in DUNE from atmospheric $\nu_\mu$ and $\bar{\nu}_\mu$ from all angles with energies above 100 MeV, with most of the events arising from neutrinos with energies below 5 GeV. Even though their flux is similar, $\bar{\nu}_\mu$ contribute less to the event rate due to relatively smaller cross-section. However, not all of these events would be considered as background in the search for DM annihilation in the Sun. The irreducible background comes from the atmospheric neutrinos that `appear' to be coming from the direction of the Sun. 

The LArTPC technology that will be employed by the DUNE detector is expected to have excellent energy and angular resolution for charged particles, especially at the energies of interest \cite{DUNE:2020ypp}. Even if the $\mu^\pm$-tracks are perfectly reconstructed, the energy and direction of the incident neutrino can only be imperfectly determined. Since our aim is to determine the direction of the flight of the neutrino in DUNE, we need to determine event-selection criteria that minimize this uncertainty. We follow Ref. \cite{Rott:2016mzs} for this purpose, albeit for high-energy neutrinos. The event selection criteria for $\mu^\pm$-tracks in DUNE has two components: a muon energy threshold ($E_\mu^{\rm th}$), and the other, on maximum value of the relative direction of the muon track to the Sun ($\theta_\mu^{\rm c}$). The remainder of this section is dedicated to determining appropriate values for these parameters.   

As evident from \Cref{fig:atm_int_nu}, a large fraction of atmospheric neutrino events can be reduced by selecting high-energy tracks above a threshold ($E_\mu^{\rm th}$). These tracks that come from the tail of the $\mu^\pm$ energy distribution. In this work, we consider three benchmark cases for $E_\mu^{\rm th}$: 500 MeV, 1 GeV, and 5 GeV. As similar cuts will be eventually imposed on the DM signal as well, one has to be conservative because very strict selection criteria will also reduce event-rate from DM annihilation, thus resulting in low sensitivity. While there may exist alternative choices that optimize the signal-to-background ratio, we do not perform this exercise. Our choices represent a low-, moderate-, and high-energy cutoff, and we shall explore their implications. 

\begin{figure}[b]
	\centering
	\includegraphics[width=0.5\linewidth]{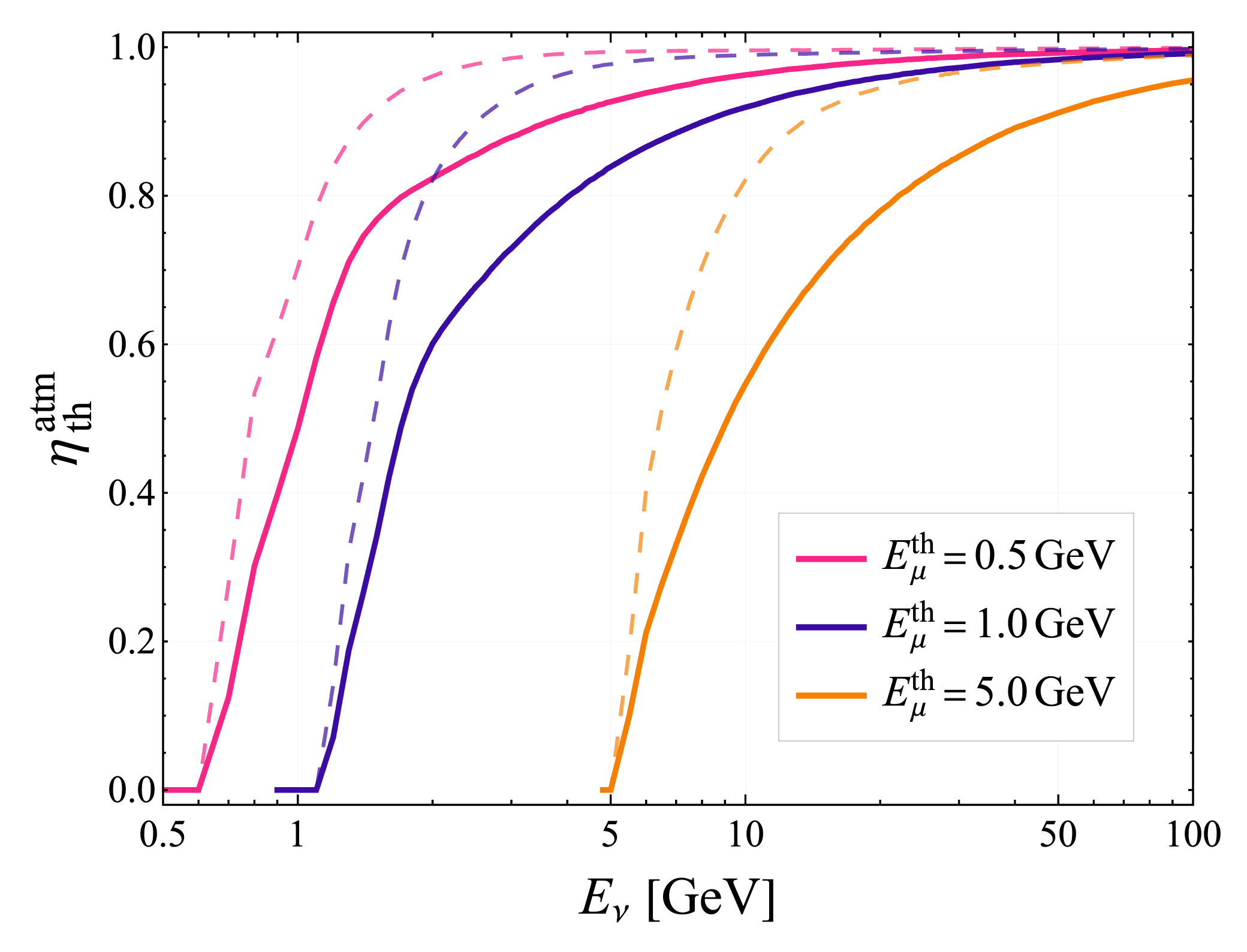}
	\caption{\label{fig:eta_th_atm} The fractions of muons produced above threshold, $\eta_{\rm th}^{\rm atm}$ (see \Cref{eq:eta_th_atm}) as a function of incident energy, for $\nu_\mu$ (solid) and $\bar{\nu}_\mu$ (dashed) are shown for three benchmark choices of the track energy threshold $E^{\rm th}_{\mu}$ as 500 MeV, 1 GeV, and 5 GeV.}
\end{figure}

The number of muon-like tracks from atmospheric neutrinos in DUNE with energy above the threshold $E_\mu^{\rm th}$ is obtained by  
\begin{equation}
	N_{\rm th}^{\rm atm} = \sum_{ \mu^\pm} \int dE_\nu \frac{dN^{\rm atm}}{dE_\nu} \times \eta_{\rm th}^{\rm atm}(E_\nu, E_\mu^{\rm th})
\end{equation}
where $\eta_{\rm th}^{\rm atm}$ is the fraction of final-state muons that are produced above threshold for a given neutrino energy, and given by, 
\begin{equation}\label{eq:eta_th_atm}
    \eta_{\rm th}^{\rm atm}(E_\nu, E^{\rm th}_{\mu}) = \frac{1}{\sigma(E_\nu)} \int\frac{d\sigma (E_\nu,E_\mu)}{dE_\mu} \Theta(E_\mu - E^{\rm th}_{\mu}) \, dE_\mu \equiv   \frac{ {\rm N_{MC}}(E_\mu > E^{\rm th}_{\mu})}{ {\rm N_{MC, \rm total}}}. 
\end{equation}
In the equation above, ${\rm N_{MC}}(E_\mu > E^{\rm th}_{\mu})$ is the number of simulated events in \texttt{NuWro} that pass the energy selection criterion, and ${\rm N_{MC, \rm total}} = 5\times10^6$ is the total number of simulated events in \texttt{NuWro} for a given neutrino/antineutrino energy. In \Cref{fig:eta_th_atm}, we show the energy dependence of $\eta_{\rm th}^{\rm atm}$ for both neutrinos and antineutrinos for the three benchmark choices of $E^{\rm th}_{\mu}$. As expected, the selected fraction of events is identically zero at low energies ($E_\nu^{\rm min} \sim m_\mu + E^{\rm th}_{\mu} $), and it monotonically increases with incident neutrino energy. We also note that the selection efficiency for anti-neutrinos is typically higher than that of neutrinos. The resulting numbers of events for muon-like tracks in DUNE from atmospheric neutrinos and anti-neutrinos are tabulated in \Cref{tab:atm_events} for the three benchmark energy threshold and assuming an exposure of 34 kton-yr.  

\begin{table}[t]
	\renewcommand{\arraystretch}{1.2}
	\centering
	\begin{tabular}{l  c  c  c }
		\toprule
		Component & $E_\mu^{\rm th}$=500 MeV & $E_\mu^{\rm th}$=1 GeV & $E_\mu^{\rm th}$=5 GeV\\
		\midrule
		$\nu_\mu$ & 884 & 552 & 120 \\ 
		$\overline{\nu}_{\mu}$ & 325 & 232 & 58 \\
		Total & 1209 & 784 & 178 \\
		\bottomrule
	\end{tabular}
	\caption{\label{tab:atm_events} The expected number of events in DUNE from atmospheric muon neutrinos and antineutrinos assuming an exposure of 34 kton-yr. The event rates are shown for the three benchmark values of the muon energy threshold, $E_\mu^{\rm th}$. We expect $\sim$2300 events/(34 kton-yr) from the neutrinos and anti-neutrinos above 100 MeV.}
\end{table}

\begin{figure}[t]
    \centering
    \includegraphics[width=0.75\textwidth]{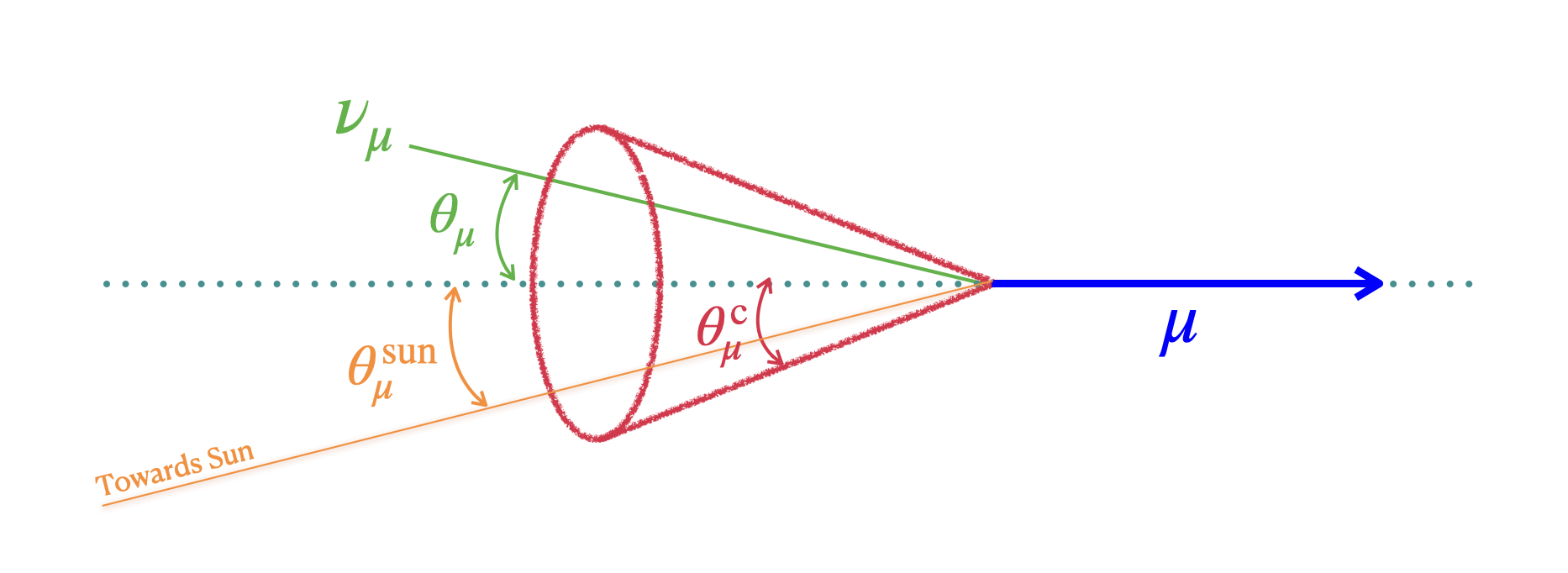}
    \caption{Geometry of a charged-current interaction is illustrated. DUNE has excellent energy and angular resolution for tracks. However, the direction of the flight of incident neutrino can only be determined statistically.}
    \label{fig:cartoon}
\end{figure}

For a given muon-like track in DUNE, one can reconstruct the initial energy of the charged particle based on the rate of energy deposition along the track. Assuming perfect resolution, one can also obtain the initial direction of the charged particle. This is, however, insufficient to infer the direction of the initial neutrino that interacted in the detector due to the lack of one-to-one correspondence between the neutrino and charged-lepton directions. One can only statistically infer whether the incident neutrino path lies within a cone of half-angle $\theta_\mu^{\rm c}$ with respect to the direction of the charged particle: $\theta_\mu<\theta_\mu^{\rm c}$ where $\theta_\mu $ is the true angle between the muon and neutrino trajectories. Depending on the position of the Sun with respect to the muon track, $\theta_{\mu}^{\rm \,sun}$, we want to determine a criterion to accept/reject the event. The geometry of the interaction is illustrated in \Cref{fig:cartoon}.   

\begin{figure}[h!]
	\centering
	\includegraphics[width=0.49\linewidth]{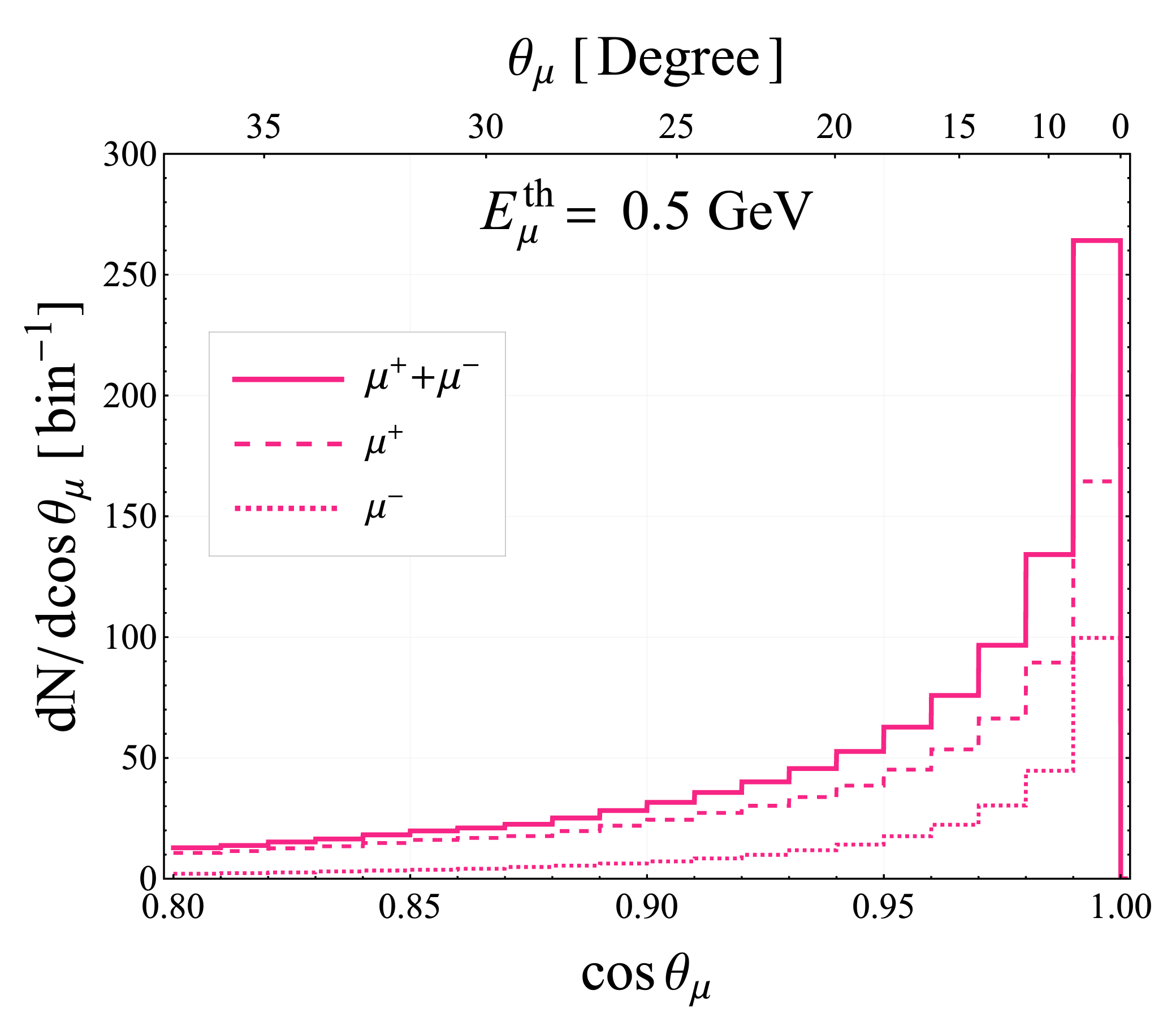}
	\includegraphics[width=0.49\linewidth]{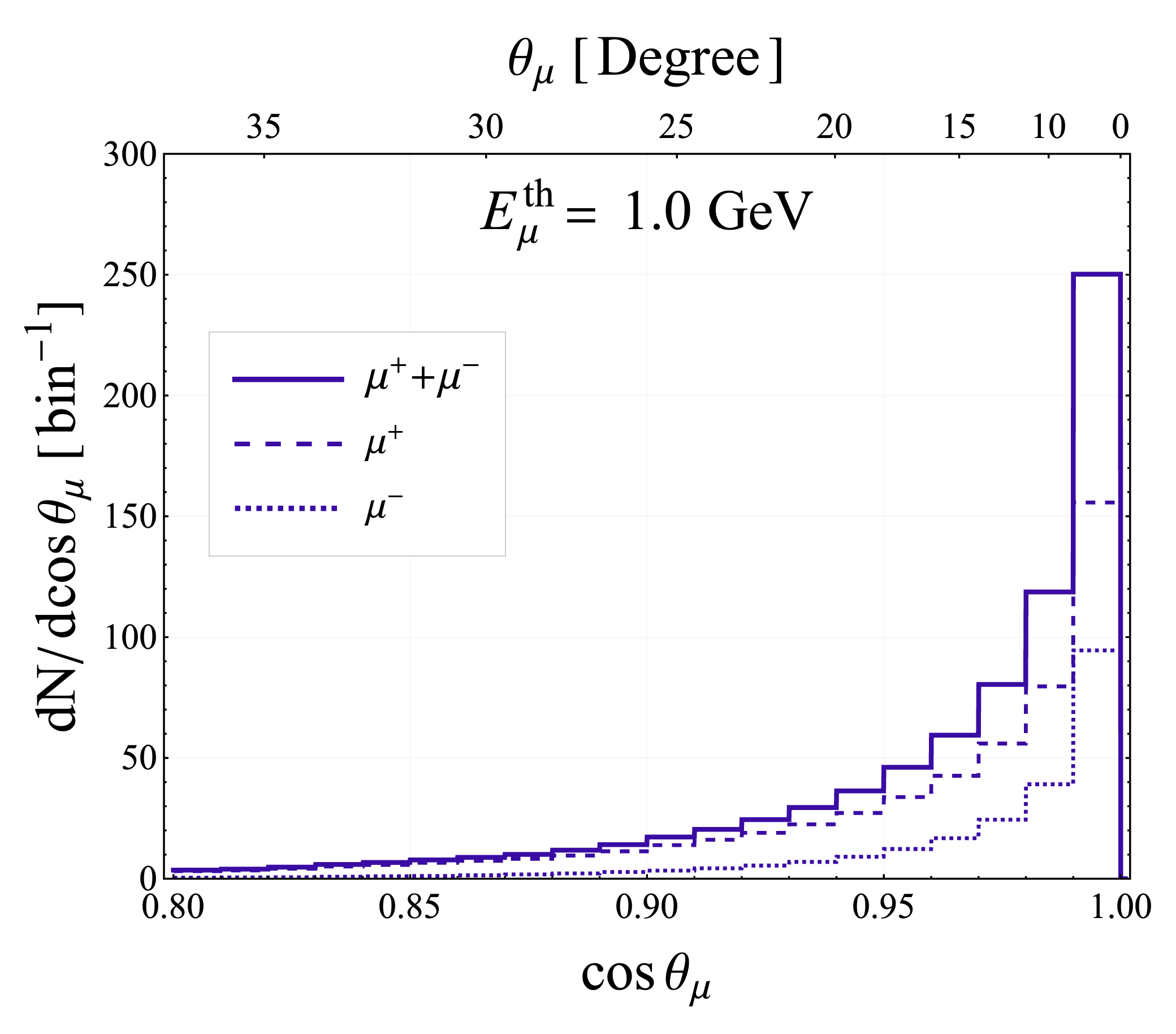}
	\includegraphics[width=0.49\linewidth]{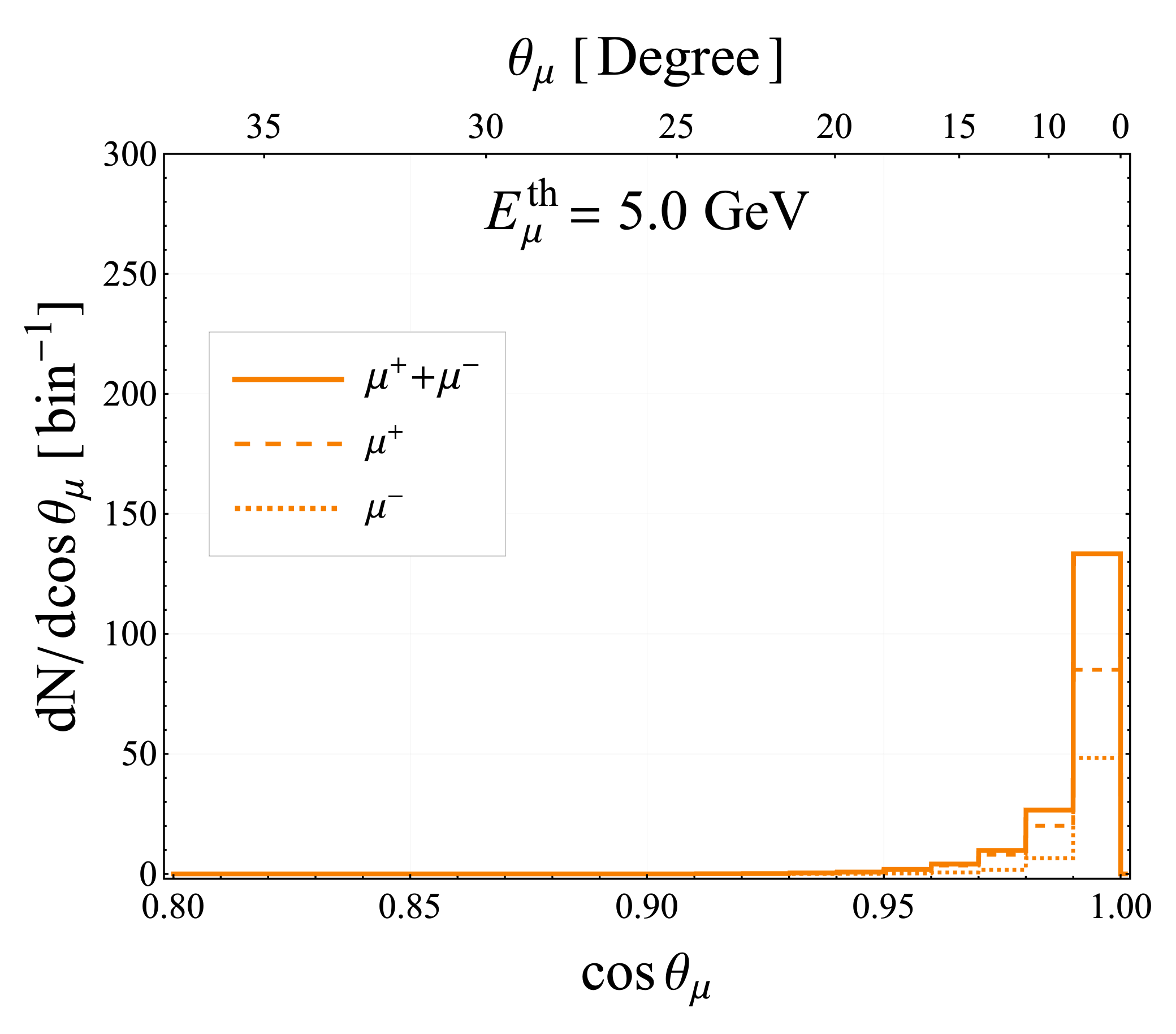}
        \includegraphics[width=0.49 \linewidth]{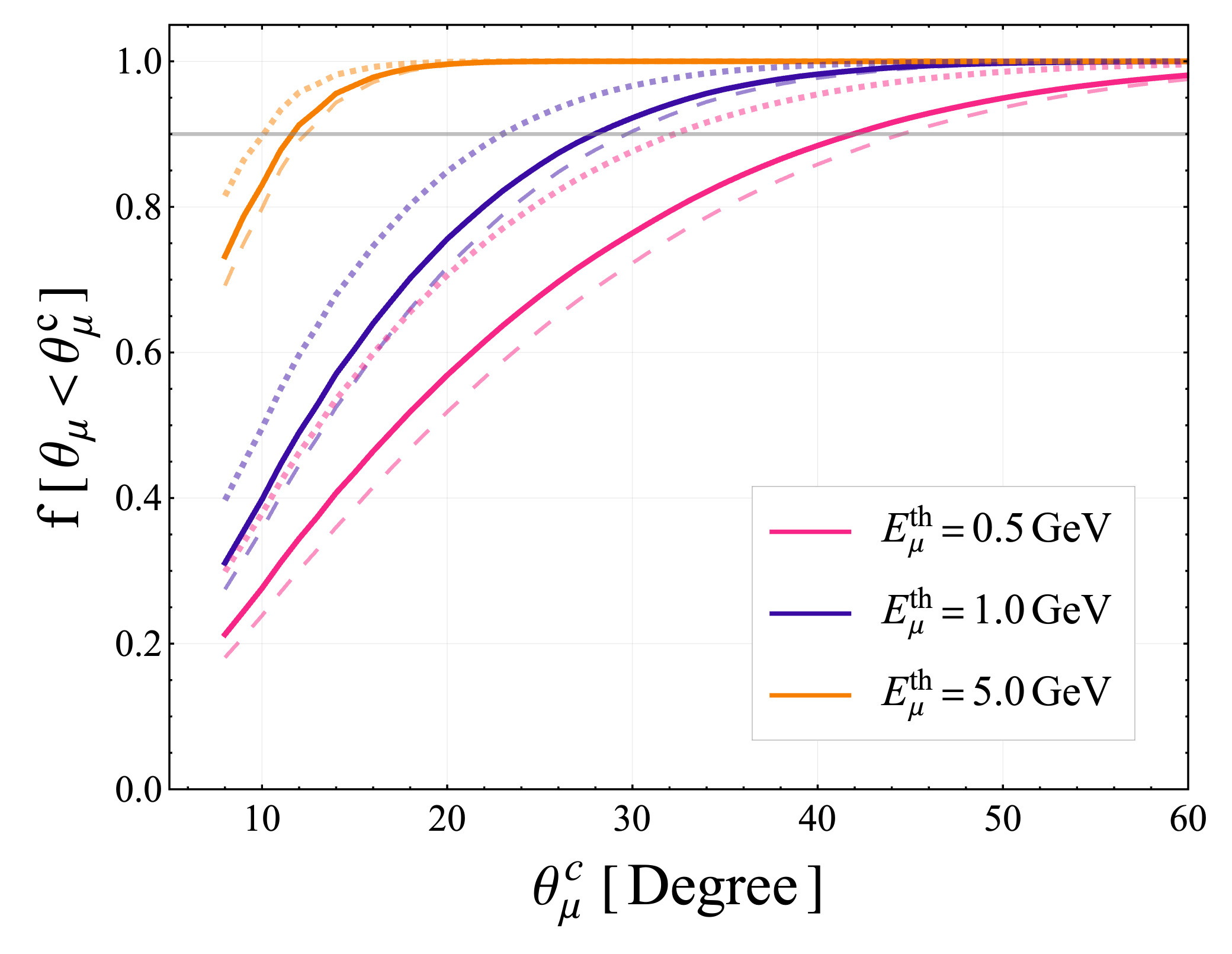}
	\caption{\label{fig:atmos_pdf_cdf} The first three figures (\emph{top left, top right, and bottom left}) show the angular distribution of muon-like events originating from atmospheric neutrinos in DUNE for the three benchmark choices of the energy threshold, $E^{\rm th}_{\mu}$, considered in this paper. We show the contributions of $\mu^-$ (dashed) and $\mu^+$ (dotted) separately, but only their aggregate (solid) is observable. \emph{Bottom right:} The fraction of events for which $\theta_\mu < \theta_\mu^{\rm \,c}$ as a function of $\theta_\mu^{\rm \,c}$ is shown for the three benchmark choices of $E^{\rm th}_{\mu}$ for $\mu^++\mu^-$ (solid), $\mu^-$ (dashed) and $\mu^+$ (dotted). The horizontal grey line is $f[\theta_\mu<\theta_\mu^{\rm \,c}]=0.90$ which determines the numerical value of $\theta_\mu^{\rm \,90}$.}
\end{figure}

To determine this directionality criterion, we look at the angular distribution of $\mu^-$ and $\mu^+$ originating from atmospheric $\nu_\mu$ and $\bar{\nu}_\mu$ respectively, as well as their aggregate. The distributions obtained for the three benchmark values of $E^{\rm th}_{\mu}$ are shown in the first three panels of Figure \ref{fig:atmos_pdf_cdf}. As expected, the muons (anti-muons) are more collinear with neutrino (anti-neutrino) direction for larger values of the muon energy threshold, and the total event rate is smaller. From this angular distribution, we determine a numerical value of $\theta_\mu^{\rm c}$ such that nearly 90\% of the incident neutrinos lie inside the cone of half-angle $\theta_\mu^{\rm c}$ with the track as axis and the interaction point as the vertex of the cone, and call it $\theta_\mu^{\rm \,90}$. The fraction of events for which $\theta_\mu < \theta_\mu^{\rm c}$ as a function of $\theta_\mu^{\rm c}$ is shown in the last panel of \Cref{fig:atmos_pdf_cdf}. Although we have explicitly shown the contributions of $\mu^-$ and $\mu^+$, only $\mu^- + \mu^+$ is observable as none of the detectors in consideration can differentiate between them. It is clear that the value of $\theta_\mu^{\rm \,90}$ depends on the choice of $E^{\rm th}_{\mu}$, and subsequently we have three benchmark scenarios:
\begin{enumerate}
	\item Low-cutoff: $E^{\rm th}_{\mu}$ = 0.5 GeV and $\theta_\mu^{\rm \,90}$ = 40$\degree$
	\item Moderate-cutoff: $E^{\rm th}_{\mu}$ = 1 GeV and $\theta_\mu^{\rm \,90}$ = 30$\degree$
	\item High-cutoff: $E^{\rm th}_{\mu}$ = 5 GeV and $\theta_\mu^{\rm \,90}$ = 15$\degree$.
\end{enumerate}
In principle, it should be possible to perform an event-by-event analysis where the muon-like track is accepted only if $E_\mu > E^{\rm th}_{\mu}$ and the position of the Sun in the sky relative to the muon track is inside the conical section with half-angle $\theta_\mu^{\rm \,90}$, i.e., $\theta_\mu^{\rm \,sun} < \theta_\mu^{\rm \,90}$. The geometry of the interaction is illustrated in \Cref{fig:cartoon}.

It is possible to improve the atmospheric neutrino background estimate by including the full angular dependence of the atmospheric neutrino flux, however, this is beyond the scope of the paper and left as future work which could also include true predictions for atmospheric neutrinos at the location of DUNE, as well as the uncertainty from reconstruction. For our estimates, we assume that the atmospheric neutrino flux at these energies is isotropic and subsequently, the event selection efficiency considering both the energy threshold and direction of the track is, 
\begin{equation}
	\eta_{\rm sel}^{\rm atm} =  \eta_{\rm th}^{\rm atm} \times \frac12 \left(1 - \cos \theta_\mu^{\rm \,90} \right)
\end{equation} 
where both $\eta_{\rm th}^{\rm atm}$ and $\theta_\mu^{\rm \,90}$ depend on the choice of $E^{\rm th}_{\mu}$. Subsequently, the \emph{irreducible} number of events from atmospheric neutrinos in DUNE is given by,
\begin{align}
    N_{\rm sel}^{\rm atm} &= \sum_{ \mu^\pm} \int dE_\nu \frac{dN^{\rm atm}}{dE_\nu} \times \eta_{\rm sel}^{\rm atm}(E_\nu, E_\mu^{\rm th}, \theta_\mu^{\rm \,90}) \\
    &\equiv N_{\rm th}^{\rm atm} \times \frac12 \left(1 - \cos \theta_\mu^{\rm \,90} \right),
\end{align}
where $N_{\rm th}^{\rm atm}$ are given in \Cref{tab:atm_events}. Any event selection criterion on the background must also be imposed on the signal, and is explored in the next section. 

Another potential source of background comes from solar atmospheric neutrinos \cite{Ng:2017aur,Arguelles:2017eao, Edsjo:2017kjk}. These are produced in cosmic ray interactions with the Sun's atmosphere. These neutrinos are typically sub-dominant as compared to Earth's atmospheric neutrinos, even when isolating events in the direction of the Sun. For DUNE, we estimate the event rate from solar atmospheric neutrinos to be $\leq 1$/(400 kton-yr). We have evaluated the neutrino ``floor" for spin-independent interactions and find that it is approximately two orders of magnitude below the projected sensitivity of shoulder neutrinos. We do not explicitly show the neutrino floor for brevity.

\section{Results and Discussion}  
\label{sec:results}

In the previous section, we identified three benchmark event-selection criteria that can minimize the background from atmospheric neutrinos, and we obtained the source pointing resolution of DUNE. In this section, we evaluate the impact of these criteria on the signal from dark matter annihilation in the Sun to obtain the projected sensitivity for DUNE from shoulder neutrinos. For the spike neutrinos, we use \Cref{eq:spike_lim} and the recent DUNE results summarised in \Cref{eq:spike_lim_dune}, together with \Cref{eq:kcap} for our new constraints inelastic dark matter. 

\begin{figure}[b]
    \centering
    \includegraphics[width=0.49 \linewidth]{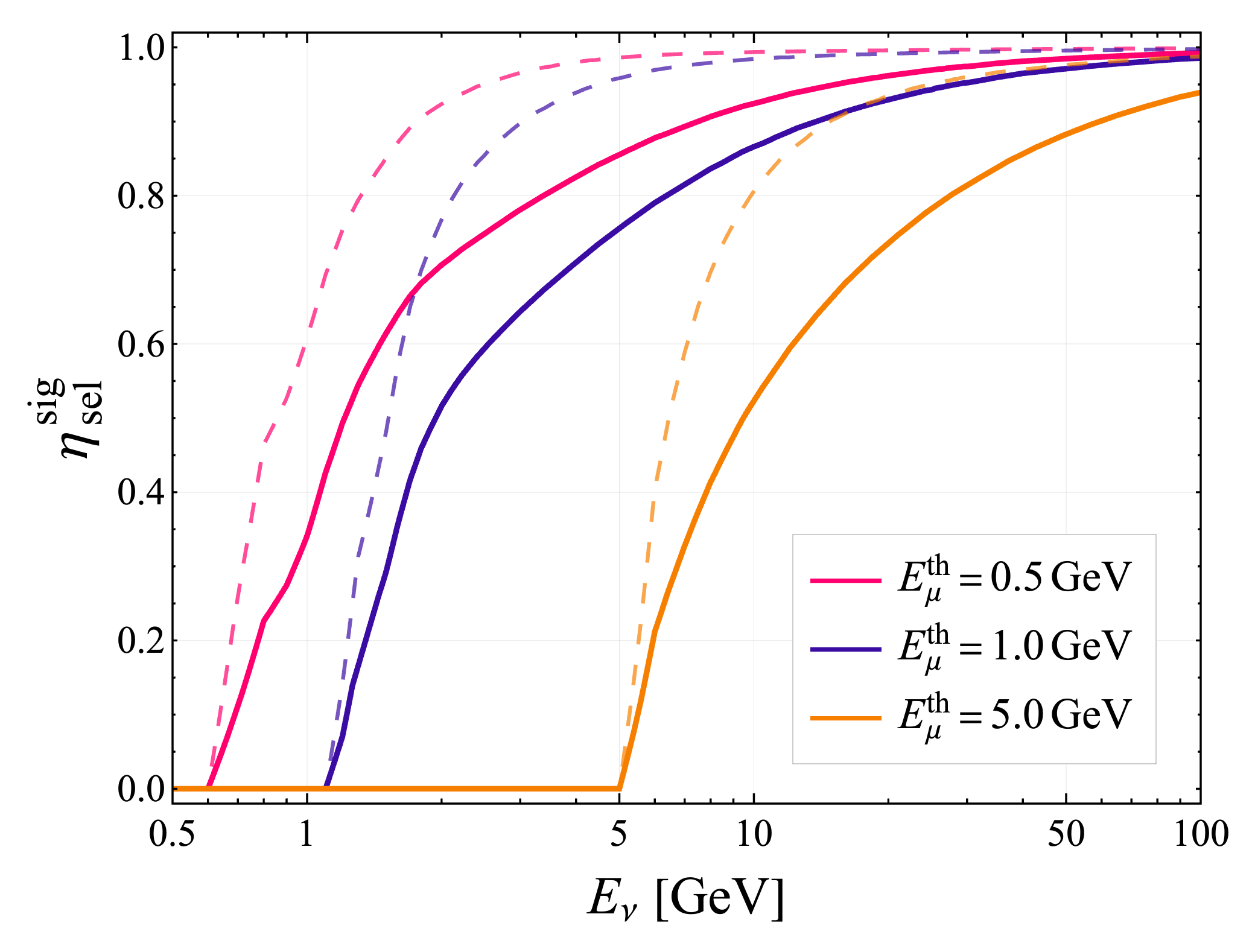}
    \includegraphics[width=0.49 \linewidth]{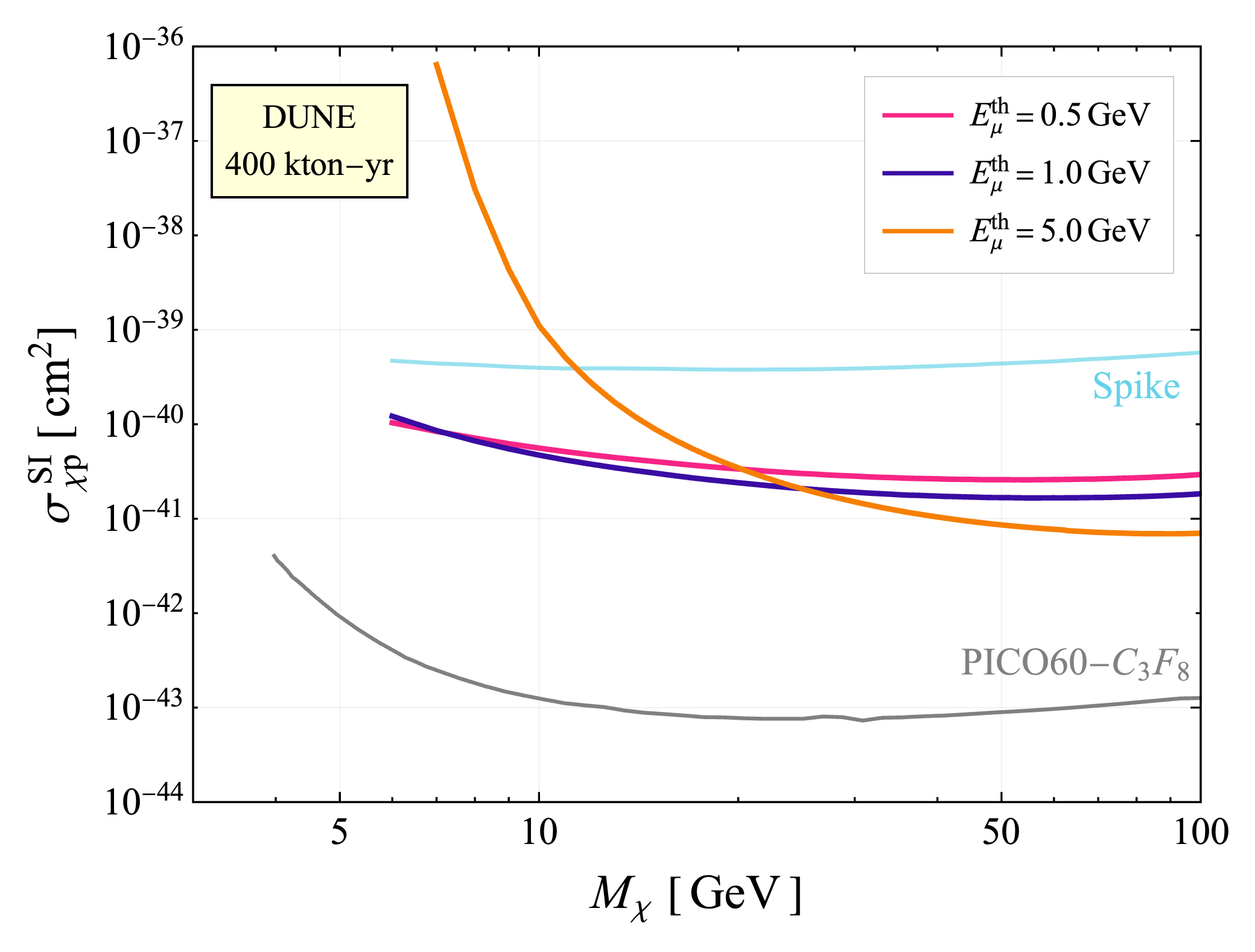}
    \caption{\label{fig:signal_sel} \textit{Left}: The signal-selection efficiency, $\eta_{\rm sel}^{\rm sig} $, for the three benchmark choices of muon energy thresholds ($E^{\rm th}_{\mu}$) are shown for $\nu_\mu$ (solid) and $\overline{\nu}_\mu$ (dashed). The cone half-angle ($\theta_\mu^{\rm c}$) is determined by the choice of $E^{\rm th}_{\mu}$ (see Table \ref{tab:nexcl}). \textit{Right}: The sensitivities of DUNE to the shoulder neutrinos from DM annihilation to $b\bar{b}$ for the three benchmark choices are shown along with the sensitivity to spike neutrinos (blue) from Ref. \cite{DUNE:2021gbm}, for elastic DM ($\delta=0)$.  The exclusion limits from the direct-detection experiment PICO60-C3F8 \cite{PICO:2019vsc} are shown using a gray curve.}
\end{figure}

Due to the lack of one-to-one correspondence between incident-neutrino energy and the outgoing-lepton energy and direction, a fraction of events from shoulder neutrinos from DM annihilation in the Sun will not satisfy the event selection criteria. A signal-event is missed if either the outgoing-lepton has kinetic energy below threshold, or if it is produced outside the cone of half-angle $\theta_\mu^{c}$ with axis along the direction of the Sun at the time of event. The fraction of the events from DM annihilation in the Sun that pass the selection criteria are given by,   
 \begin{equation}
	N_{\rm sel}^{\rm sig} = \frac{M_{\rm target }\, T}{\text{kton}} \frac{N_A \times 10^9}{A} \times  \sum_{ \mu^\pm} \int dE_\nu\,  \sigma_{\rm tot}(E_\nu) \, \frac{d\Phi^{\rm D}}{dE_\nu} \times \eta_{\rm sel}^{\rm sig}(E_\nu, E_\mu^{\rm th}, \theta_\mu^{\rm \,90} )
\end{equation}
where $\eta_{\rm sel}^{\rm sig}$ is the \emph{signal-selection efficiency}, and determined by,
\begin{align}
	\label{eq:eta_sel_sig}
	\eta_{\rm sel}^{\rm sig}(E_\nu, E^{\rm th}_{\mu}, \theta_\mu^{\rm \,90}) &= \frac{1}{\sigma_{\rm tot}(E_\nu)} \int {dE_\mu\, d\cos \theta_\mu}\frac{d\sigma}{dE_\mu d\cos \theta_\mu}\, \Theta(E_\mu - E^{\rm th}_{\mu}) \,\Theta(\theta_\mu^{\rm \,90}-\theta_\mu)\,\nonumber\\
	&\equiv   \frac{ {\rm N_{MC}}(E_\mu > E^{\rm th}_{\mu}, \theta_\mu < \theta_\mu^{\rm \,90})}{ {\rm N_{MC, \rm total}}}. 
\end{align}
We have evaluated $\eta_{\rm sel}^{\rm sig}$ for $\nu_\mu$ and $\bar{\nu}_\mu$ using \texttt{NuWro} for the three benchmark scenarios considered in this paper, and the results are shown in Figure \ref{fig:signal_sel} (left). 

\begin{table}[t]
	\renewcommand{\arraystretch}{1.2}
	\centering
	\begin{tabular}{c c c c}
		\toprule 
		$E^{\rm th}_{\mu}$ & $\theta_\mu^{\rm \,90}$ & $ N^{\rm atm}_{\rm sel}$ & $ N^{\rm excl}_{90}$ \\
		\midrule 
		0.5 GeV & 40$\degree$ & 1664 & 57.5 \\ 
		1.0 GeV & 30$\degree$ & 618 & 34.4 \\
		5.0 GeV & 15$\degree$ & 36 & 9.4 \\
		\bottomrule
	\end{tabular}
	\caption{ \label{tab:nexcl} The number of expected atmospheric background events that pass the selection criterion ($N^{\rm atm}_{\rm sel}$) and the number of signal events needed for 90\% C.L. exclusion ($N^{\rm excl}_{90}$) are tabulated. We have assumed an exposure of 400 kton-yr for DUNE.}
\end{table}
To obtain the projected 90\% C.L. exclusion limits, we follow Ref. \cite{Rott:2016mzs} and assume a Poisson distribution for the event-rate from the atmospheric background. We assume a detector exposure of 400 kton-yr, and the observed number of events is set equal to the expected number of background events, rounded to  the nearest integer. A signal is considered excluded at 90\% C.L., if the total number of events, i.e., the sum of background and signal, exceeds the observed number of events in at least 90\% of the cases. For the three scenarios considered in this paper, the number of expected atmospheric background events that pass the selection criterion ($N^{\rm atm}_{\rm sel}$), and the number of signal events needed for 90\% C.L. exclusion ($N^{\rm excl}_{90}$) are presented in Table \ref{tab:nexcl}. 

The projected sensitivity in the $M_\chi - \sigma_{\chi p}^{\rm SI}$ plane for elastic DM (i.e., $\delta = 0$) is obtained by considering $N_{\rm sel}^{\rm sig} \geq N_{90}^{\rm excl{\color{white}g}}$. The results are shown in Figure \ref{fig:signal_sel} (right). It is interesting to note that the low-energy-cutoff scenario has larger sensitivity at small DM mass ($\leq$ 7 GeV) and the high-energy-cutoff scenario has larger sensitivity at larger DM mass ($\geq$ 20 GeV). This can be attributed to two different factors. First, a larger threshold on the muon energy implies that event-rate from DM with small mass is suppressed. Second, the muons with larger energy are more collinear with the incident neutrino. As a result, not only the background is suppressed due to small $\theta_\mu^{\rm c}$, but also the fraction of events that satisfy the directionality criterion is larger, and subsequently the signal-to-background ratio is enhanced. 

Unsurprisingly, the entire parameter space for elastic DM that can lead to detectable signal in DUNE is already ruled out by direct-detection experiments. In the Figure \ref{fig:signal_sel} (right), we only show the limits by PICO60-\C3F8 from Ref. \cite{PICO:2019vsc}. Other direct-detection experiments such as XENON, LZ, and Panda-X are even more constraining. Their upper bounds on the spin-independent cross section $\sigma_{\chi p}^{\rm SI}$ are below the scale shown on this plot. This is one of the primary reasons for investigating inelastic DM, for which the direct-detection experiments are less sensitive. Moving forward, we choose the moderate-cutoff scenario (i.e., $E^{\rm th}_{\mu}$ = 1 GeV and $\theta_\mu^{\rm c}$ = 30$\degree$) to forecast our projected sensitivity for DUNE to detect the shoulder neutrinos from annihilation of inelastic DM captured in the Sun. 

\begin{figure}[t]
	\centering
 	\includegraphics*[width = 0.49\textwidth]{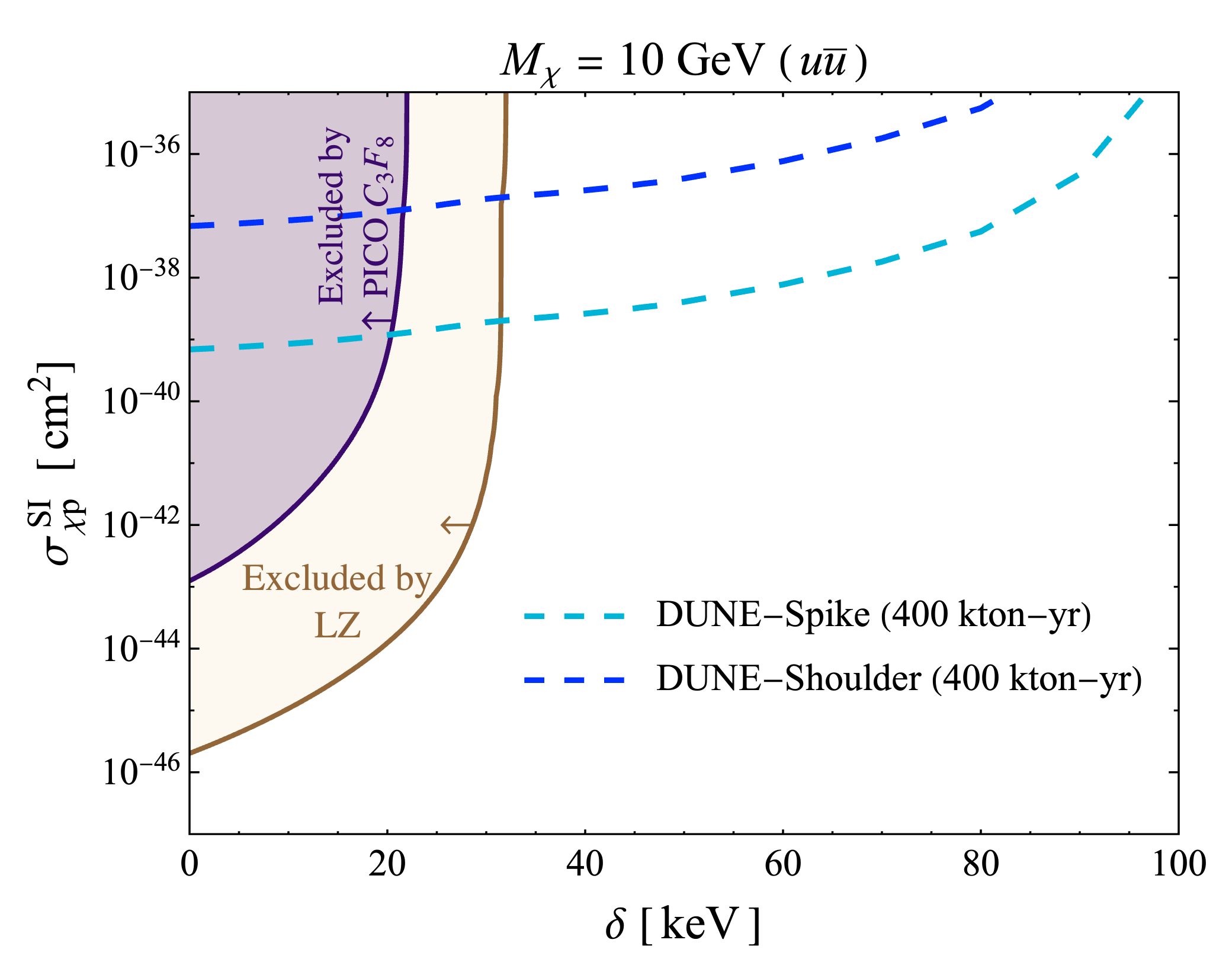}
	\includegraphics*[width = 0.49\textwidth]{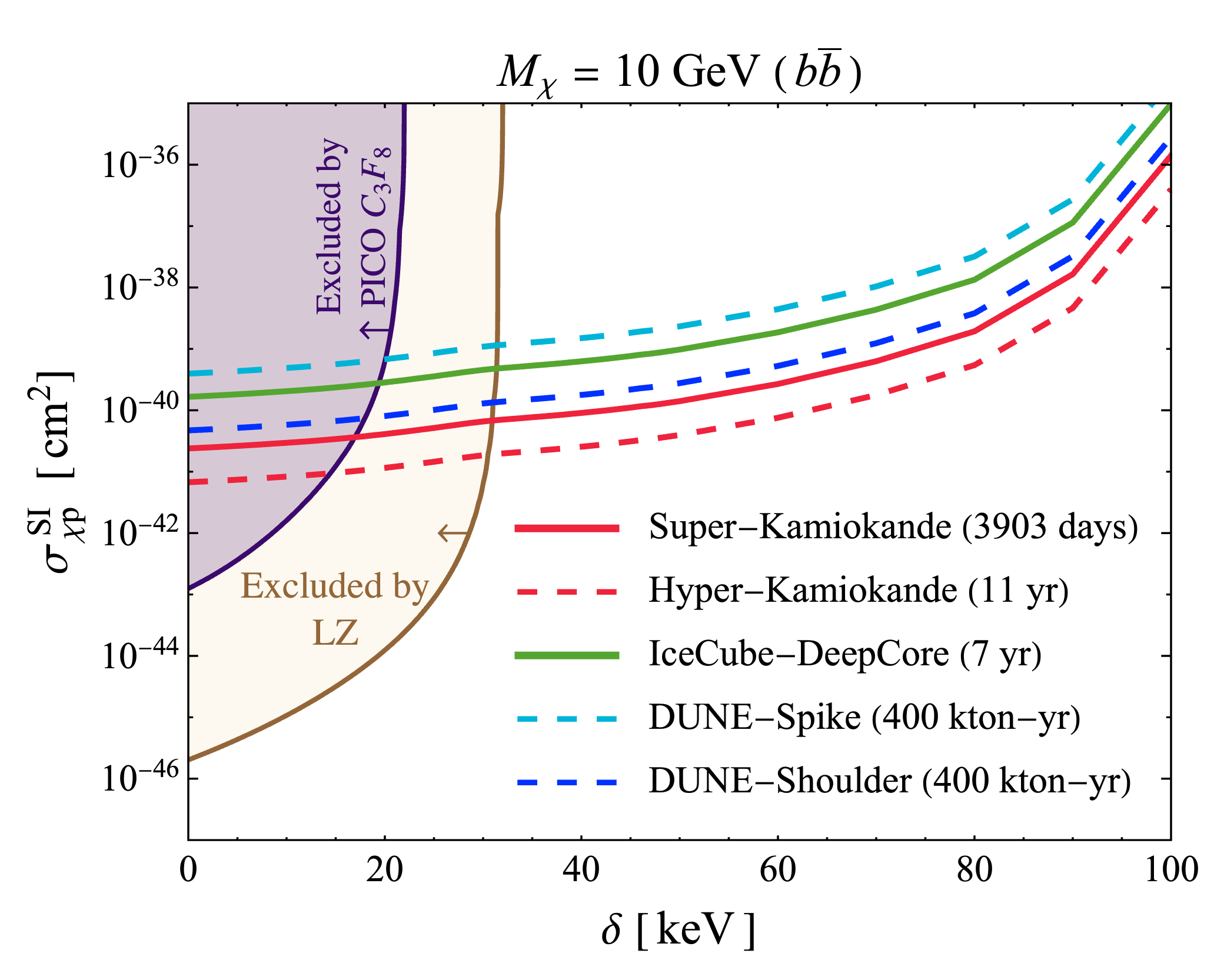}
	\includegraphics*[width = 0.49\textwidth]{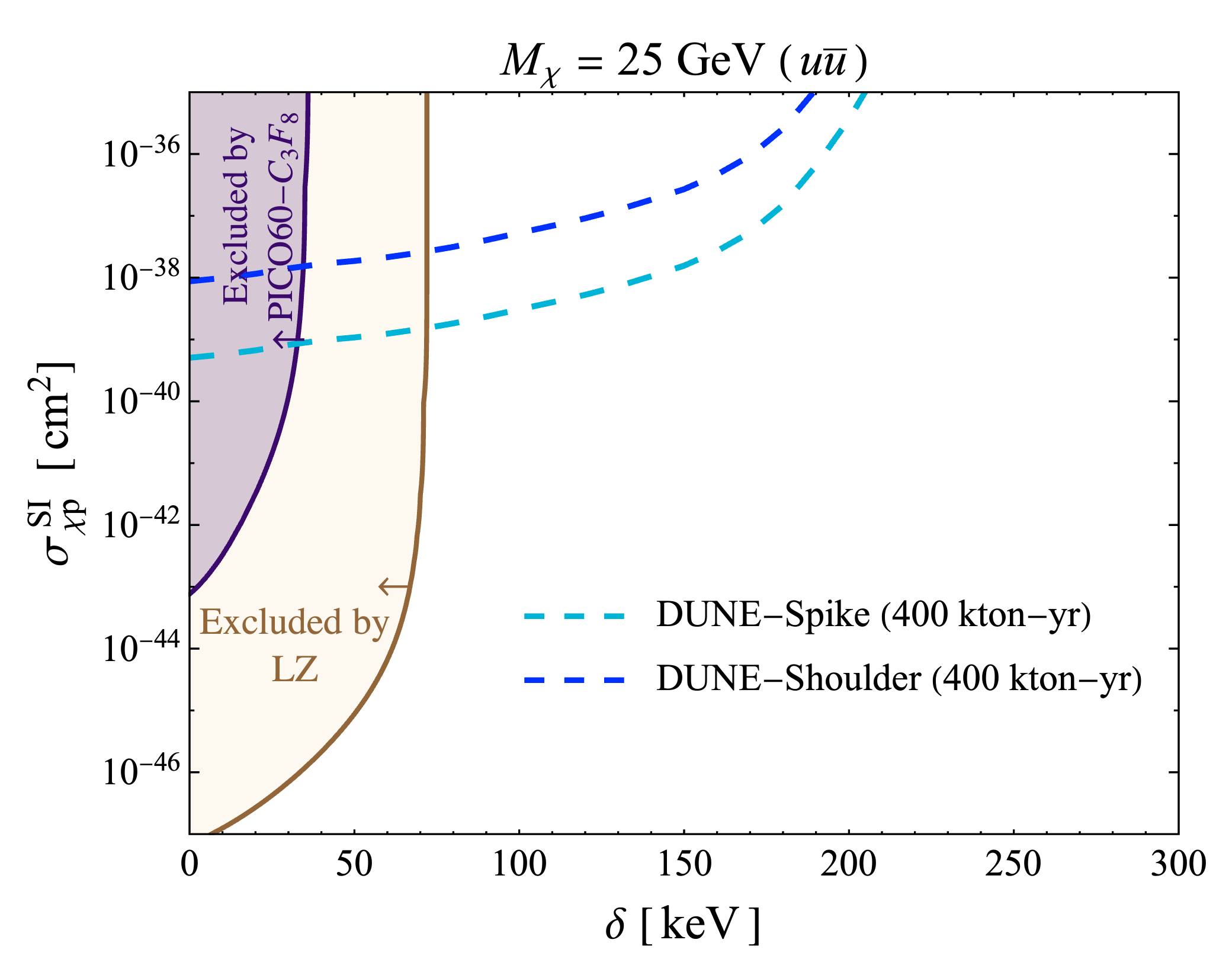}
	\includegraphics*[width = 0.49\textwidth]{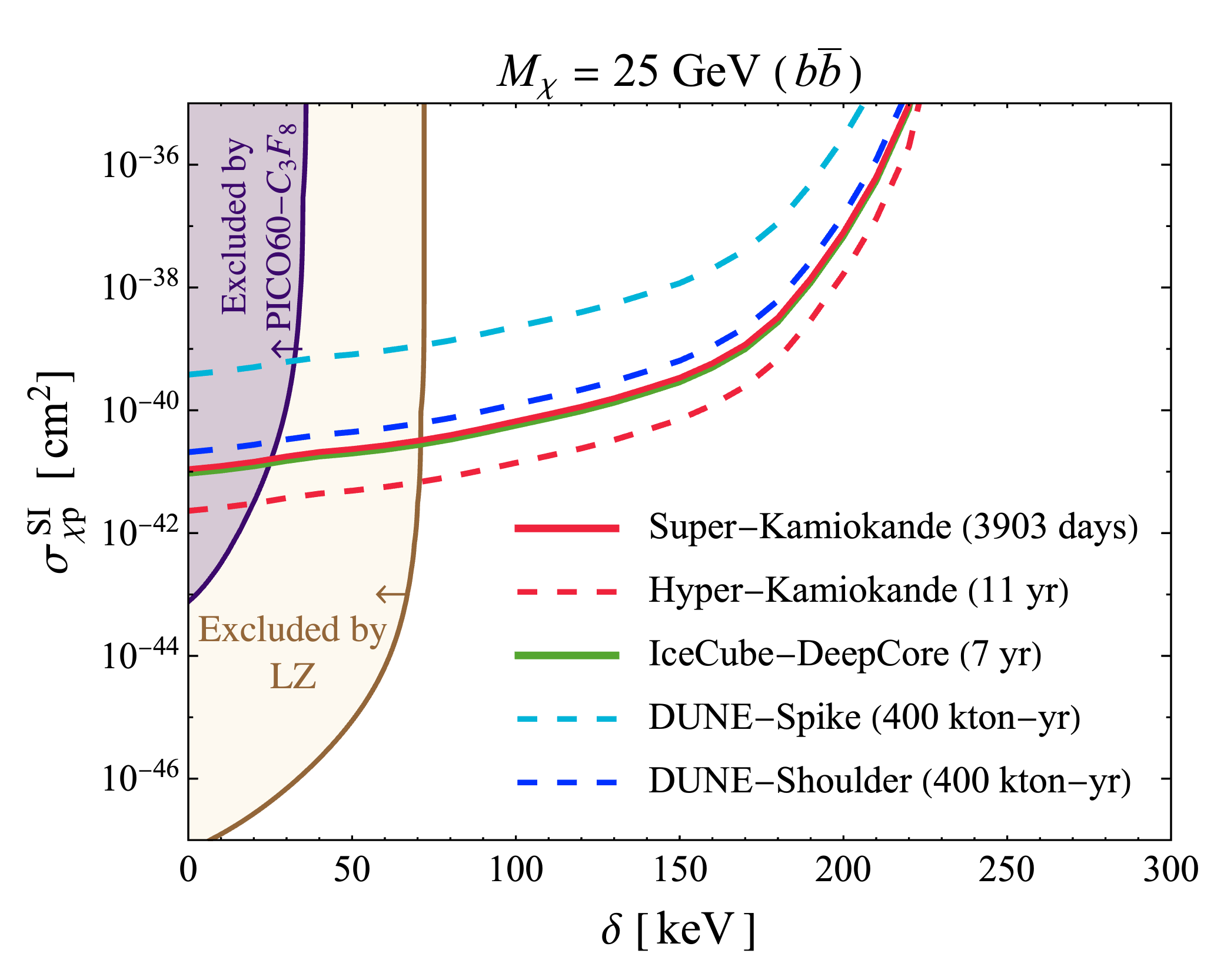}
	\caption{\label{fig:main_sigma-delta} The exclusion limits with neutrino experiments and direct-detection experiments on the parameter space of mass-splitting ($\delta$) and spin-independent scattering cross-section with nucleon ($\sigma_{\chi p}^{\rm SI}$) are shown for two choices of DM mass ($M_\chi$). The limits for DM annihilation to light-quark channel (left) and heavy-quark channel (right) are shown separately. The projected 90\% C.L. exclusion limits from DUNE with 400 kton-yr exposure are shown for spike and shoulder neutrinos independently. The existing 90\% C.L. exclusion limits from Super-Kamiokande with 3903 days exposure are shown using the red solid curve (using \Cref{eq:kcap} and limits on elastic interactions from Ref. \cite{Super-Kamiokande:2015xms}). The projected sensitivity of Hyper-Kamiokande with 11 years exposure is shown with the red dashed curve (using \Cref{eq:kcap} and Ref. \cite{Bell:2021esh}). The existing limits from IceCube-DeepCore with 7 years of exposure is shown in green (using \Cref{eq:kcap} and Ref. \cite{IceCube:2021xzo}), but note that they are weaker than Super-K at these DM masses. The limits from Super-Kamiokande and Hyper-Kamiokande are dominated by upward-muons and only available for DM annihilation to $b\bar{b}$. The shaded region is ruled out from the direct detection experiments LZ \cite{LUX-ZEPLIN:2022xrq} PICO-60 \cite{PICO:2019vsc}.}
\end{figure}

For inelastic DM with non-zero mass splitting $\delta$, the capture rate in the Sun is reduced by a factor $k_{\rm cap}(M_\chi, \delta) \leq 1$ (see \Cref{eq:kcap}) whereas the spectrum of neutrinos is not affected\footnote{If the heavier state is long-lived and constitutes a minor component of the cosmological DM density, only the capture rate in the Sun needs to be suitably modified and the spectrum of neutrinos remains the same.}. As a result, the limiting sensitivity for DM-nucleon cross-section for inelastic DM can be obtained from the limits obtained for elastic DM by a simple translation, 
\begin{equation}\label{eq:siglim}
	\sigma_{\rm cap}^{\rm lim}(M_\chi, \delta) = \sigma_{\rm cap}^{\rm lim}(M_\chi, \delta=0)\times k_{\rm cap}^{-1}(M_\chi, \delta) 
\end{equation}
where $\sigma_{\rm cap}^{\rm lim}(M_\chi, \delta=0)$ are the limits on the DM-nucleon cross-section obtained from the capture of elastic DM (shown in Figure \ref{fig:signal_sel}). In Ref. \cite{Super-Kamiokande:2015xms}, Super-Kamiokande utilized their upward-going muons data to look for neutrinos from DM annihilation in the Sun. Their limits are presented for elastic DM. We translate their $\delta=0$ limits to limits on inelastic DM using \Cref{eq:siglim}. In Ref. \cite{Bell:2021esh}, the sensitivity of Hyper-Kamiokande using upward-going muons was presented for SD interactions. We translate these limits for inelastic dark matter with an additional factor of $\Gamma_{\rm cap}^{\rm SD}(M_\chi)/\Gamma_{\rm cap}^{\rm SI}(M_\chi)$ in \Cref{eq:siglim} to account for the difference in capture rate. Analogously, the limiting sensitivity of direct-detection experiments for inelastic DM can be obtained from the limits on elastic DM,    
\begin{equation}
	\sigma_{\rm DD}^{\rm lim}(M_\chi, \delta) = \sigma_{\rm DD}^{\rm lim}(M_\chi, \delta=0)\times k_{\rm DD}^{-1}(M_\chi, \delta)
\end{equation}
where `DD' in the subscript refers to the specific direct-detection experiment. As described earlier, we limit our attention to PICO-C3F8, LZ, and PandaX for a good coverage of the range of DM mass. Moreover, the capture of inelastic DM in the Sun with spin-dependent (SD) interactions is highly suppressed \cite{Menon:2009qj, Blennow:2015hzp}, and we only present our sensitivity forecasts for spin-independent (SI) scattering cross-section of DM with nucleons. 

\begin{figure}[t]
	\centering
 	\includegraphics*[width = 0.49\textwidth]{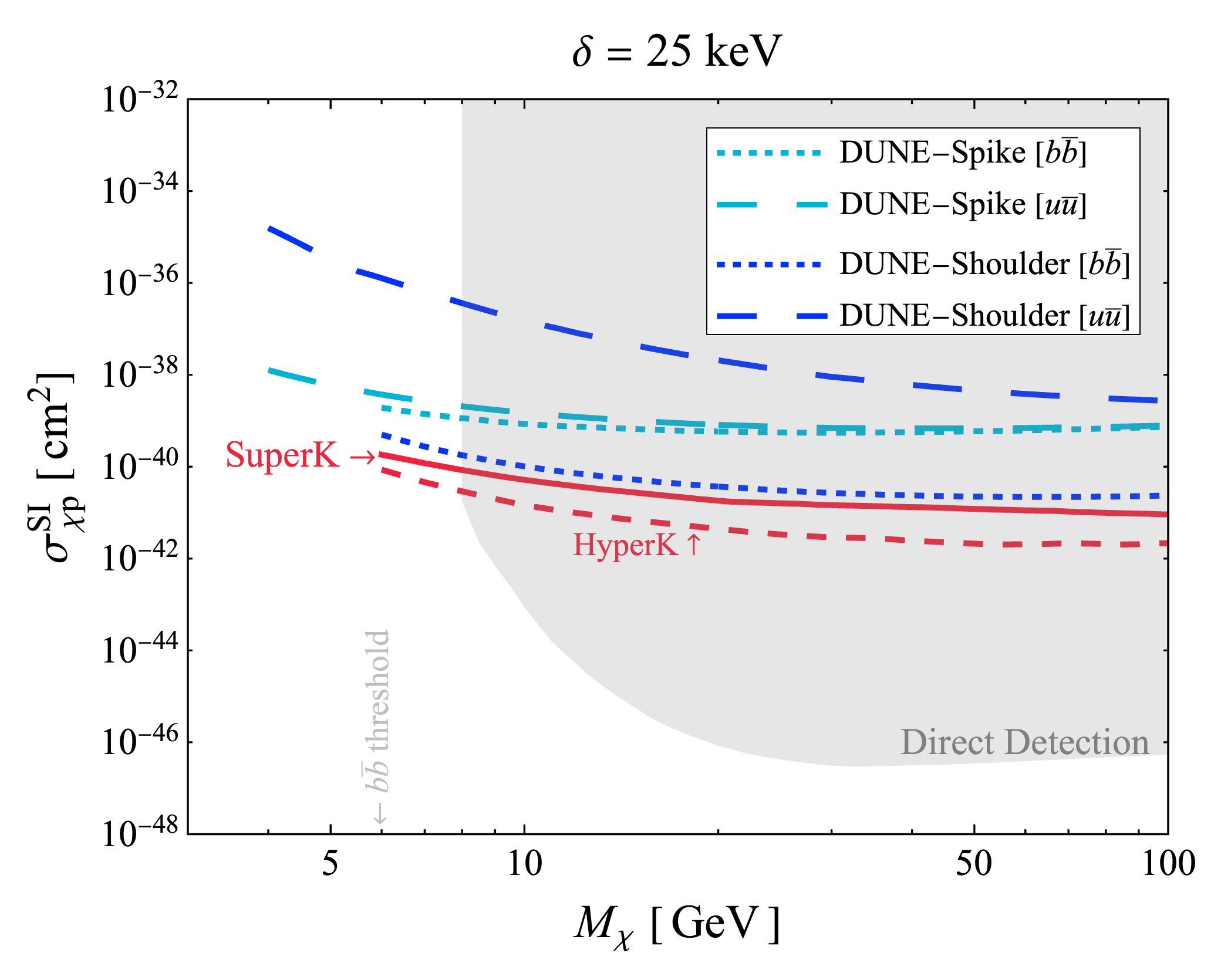}
        \includegraphics*[width = 0.49\textwidth]{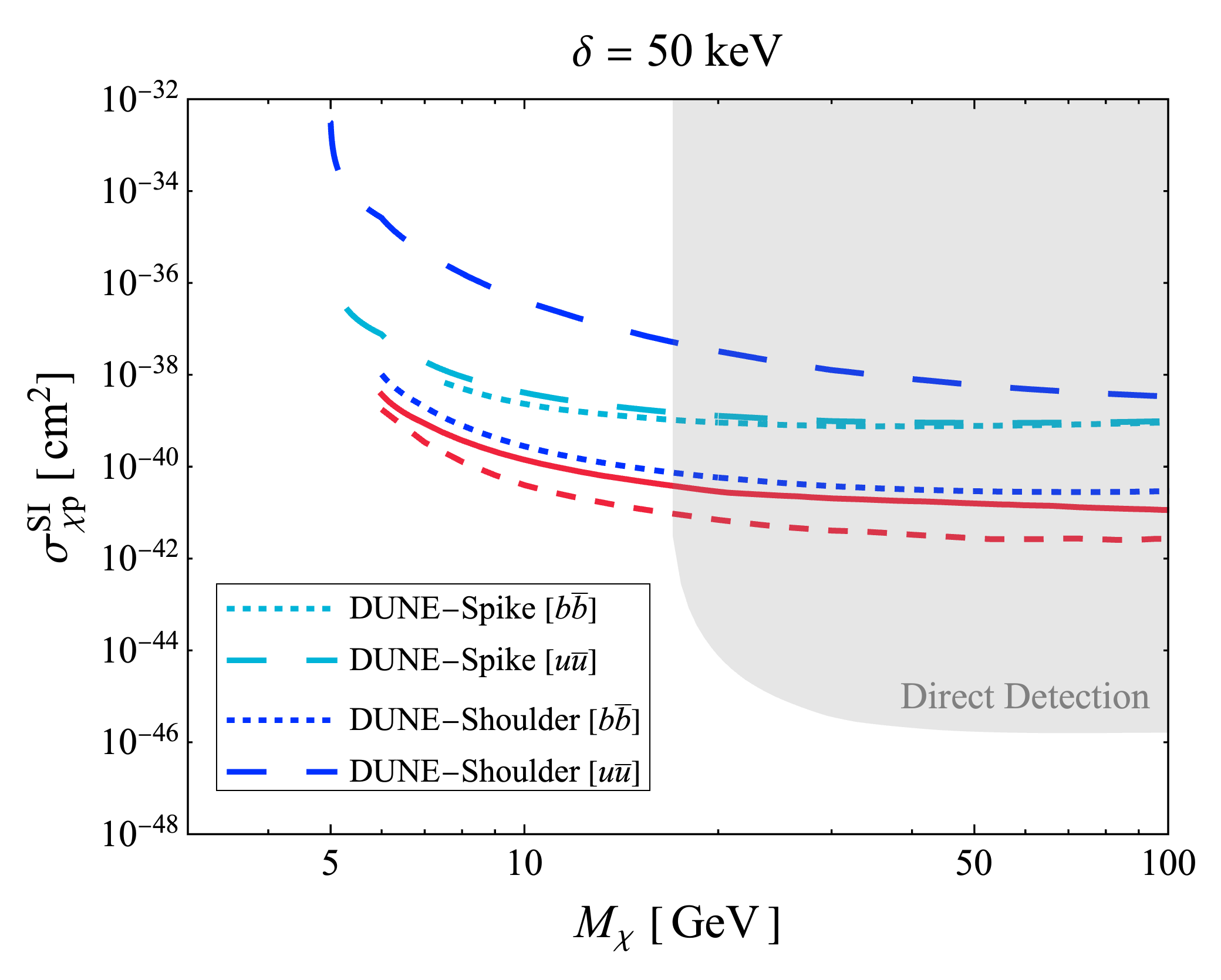}
        \includegraphics*[width = 0.49\textwidth]{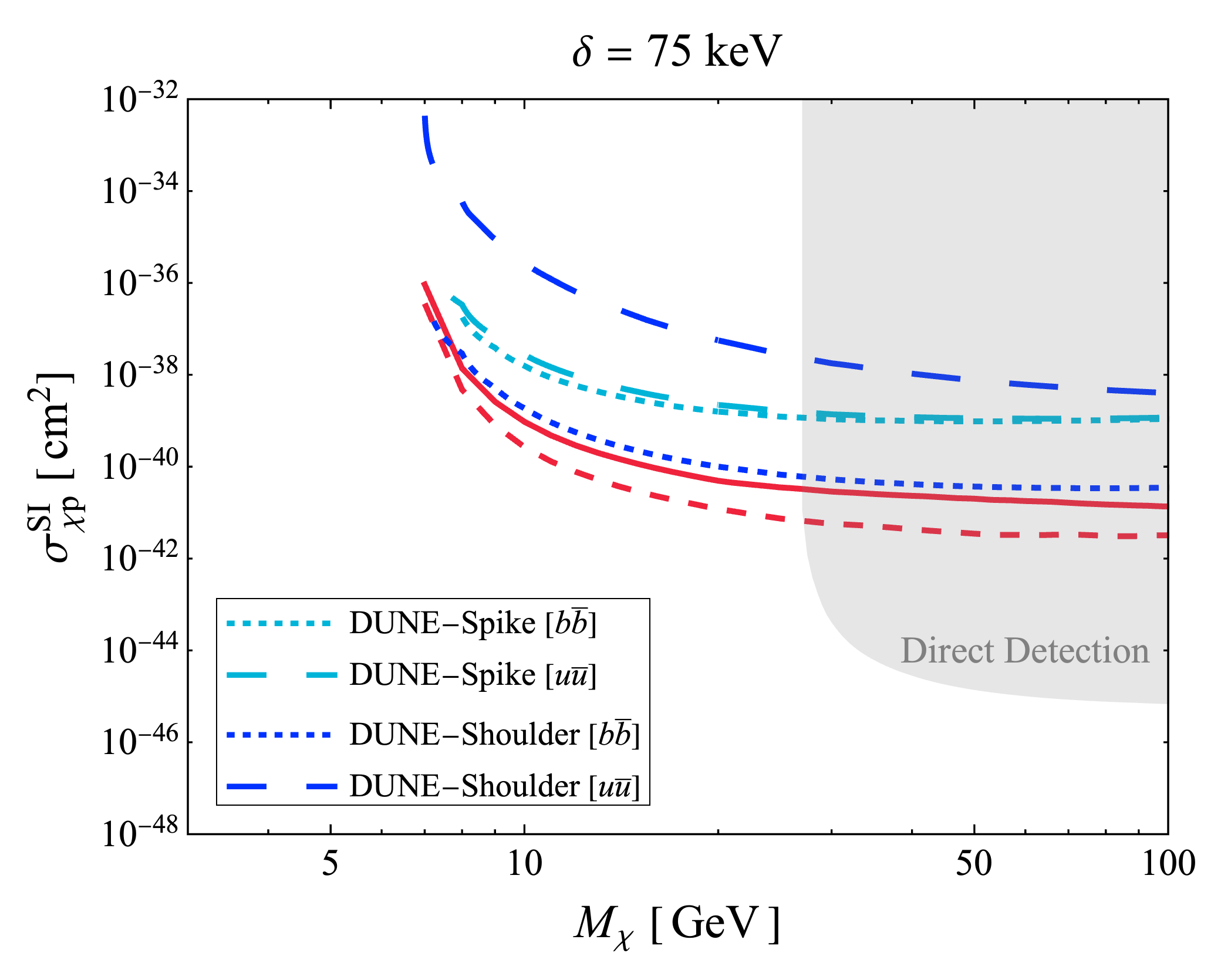}
        \includegraphics*[width = 0.49\textwidth]{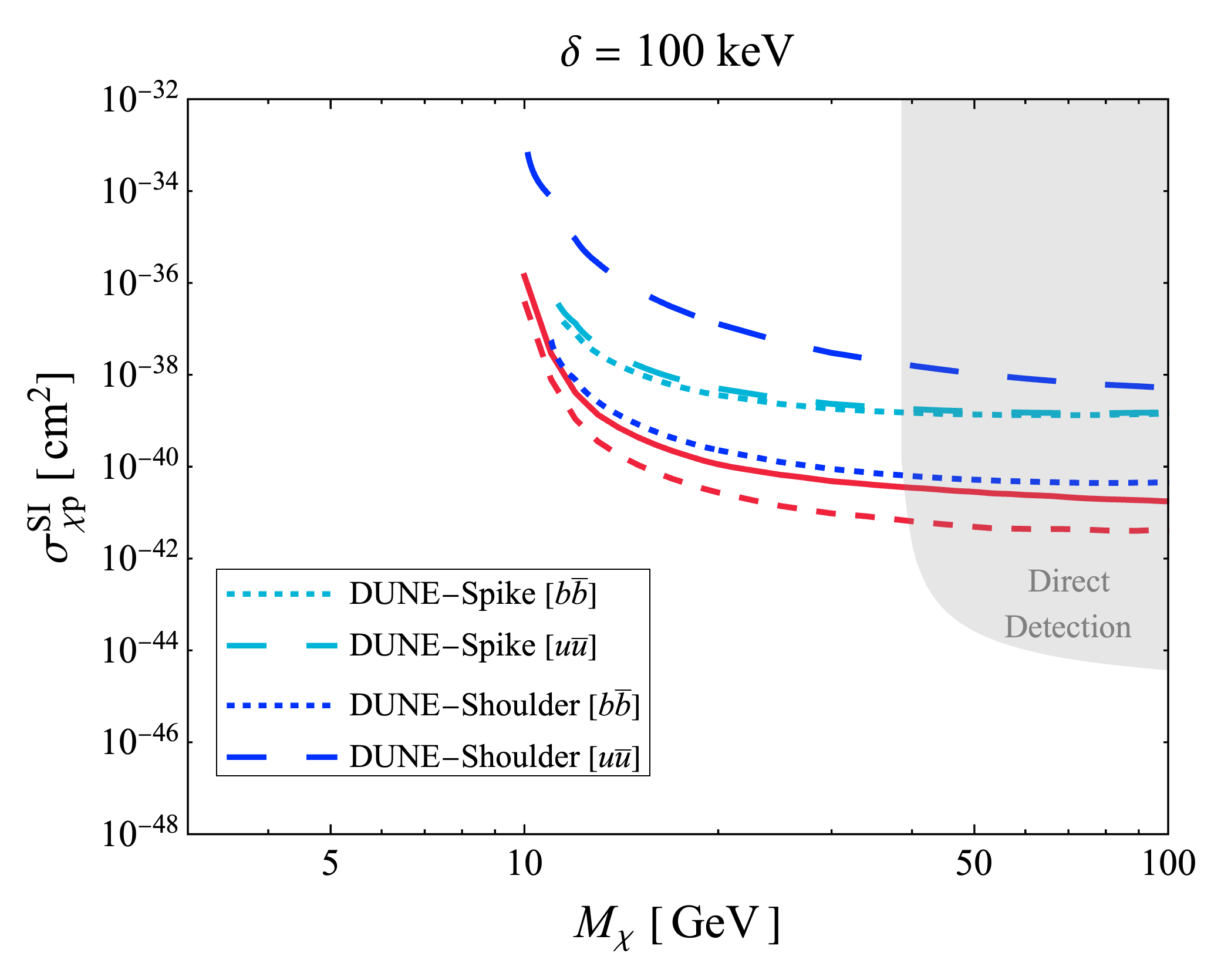}
	\caption{\label{fig:main_sigma-m} The complementarity of neutrino experiments and direct-detection experiments on the parameter space of DM mass ($M_\chi$) and spin-independent scattering cross-section with nucleon ($\sigma_{\chi p}^{\rm SI}$) is shown for four choices of mass-splitting ($\delta$). The projected 90\% C.L. exclusion limits from DUNE with 400 kton-yr exposure are shown for spike and shoulder neutrinos separately and assuming light- and heavy-quark annihilation channels independently. The existing limits Super-Kamiokande and the projected sensitivity of Hyper-Kamiokande are obtained using \Cref{eq:kcap} and Refs. \cite{Super-Kamiokande:2015xms} and \cite{Bell:2021esh} respectively. Note that the limits from Super-K and Hyper-K assumes annihilation to $b\bar{b}$ channel. The gray-shaded region is ruled out from the direct detection experiments. The existing limits from IceCube-DeepCore \cite{IceCube:2021xzo}are  not shown because they are weaker than Super-Kamiokande at lower DM mass and weaker than direct-detection at larger DM mass.}
\end{figure}

With the translated limits from other neutrino experiments and direct-detection experiments, we present our results for DUNE in \cref{fig:main_sigma-delta} and \cref{fig:main_sigma-m}. The sensitivity is projected for muon-like \emph{contained} tracks with moderate cutoff (i.e., $E_\mu^{\rm th} = 0.5$ GeV and $\theta_\mu^{\,90} = 30$\degree) and assuming a fiducial exposure of 400 kton-yr. 

In \cref{fig:main_sigma-delta}, we show our results for the light-quark channel and the heavy-quark channel separately. This is to highlight the fact that other neutrino experiments (such as Super-Kamiokande and IceCube) can only detect shoulder muon neutrinos and antineutrinos, which is prominent for heavy-quark channel only. In case of dark matter annihilation to light-quarks, the shoulder is suppressed. The $\nu_\mu$-spike is only detectable at DUNE. In principle, Super-Kamiokande and DUNE can also detect the $\nu_e$-spike (originating due to flavor conversion) for the light-quark channel, and the projected sensitivity has been obtained in \cite{Rott:2015nma}. We do not show these limits as our discussion is focused on detection of muon-like tracks. One also expects some events from tau neutrinos and anti-neutrinos from DM annihilation without any background from atmospheric neutrinos. However, their source-pointing resolution is difficult to determine without dedicated detector simulations, and we encourage DUNE collaboration to consider this. In \cref{fig:main_sigma-m}, we show the limits and projected sensitivities of neutrino experiments and direct detection experiments for four benchmark values of the mass splitting $\delta$. For brevity, we show the limits from light-quark channel and the heavy-quark channel in the same figure, even though they can be mutually exclusive. 

The main result of our new analysis of the current neutrino experiments, Super-Kamiokande and IceCube, is that they already rule out a large portion of the inelastic dark matter parameter space that is usually inaccessible to direct-detection experiments. We find that for the heavy-quark channel, the projected sensitivities from both spike and shoulder muon neutrinos at DUNE are weaker than the Super-Kamiokande limits we have determined here. Note that the limits from Super-Kamiokande are dominated by upward going muons which originate outside the detector volume, and thus have a much larger target mass, resulting in better sensitivity. In the heavy-quark channel, only Hyper-Kamiokande is expected to improve the limits in future. For the light-quark channel, DUNE can detect the $\nu_\mu$-spike efficiently. In the light-quark channel, the projected limits from shoulder neutrinos is weaker than the spike, as expected. As \Cref{fig:main_sigma-delta} and \Cref{fig:main_sigma-m} show, DUNE can constrain a region of $(M_\chi,\delta)$ parameter space for inelastic dark matter annihilation to light quarks where Super-Kamiokande, Hyper-Kamiokande and DD experiments do not have sensitivity.

\section{Summary and Outlook}
\label{sec:conclusions}

In models of dark matter with sizable interactions with standard model particles, the dark matter in the galactic halo can be gravitationally captured by the Sun. The subsequent annihilation of the captured dark matter results in a novel neutrino flux on Earth. The detectable neutrino spectrum has two components: a spike at $\sim$236 MeV from kaon decays-at-rest, and a broad-spectrum shoulder from prompt decays of mesons. DUNE is a good candidate to study the spike neutrinos, and has been widely discussed in the literature \cite{Rott:2015nma, Rott:2016mzs, Rott:2017weo, DUNE:2021gbm}, whereas the sensitivity to shoulder neutrinos is only discussed for electron neutrinos \cite{Bueno:2004dv}. We evaluate, for the first time, the sensitivity of DUNE from \emph{contained} GeV-scale tracks originating from the shoulder $\nu_\mu$ and $\bar{\nu}_\mu$. To reduce the atmospheric neutrino background, we determine event selection criteria in DUNE that can identify neutrinos coming from the direction of the Sun. Unsurprisingly, the parameter space probed by DUNE for dark matter with elastic interactions with nucleons is already ruled out by direct detection experiments. For inelastic dark matter \cite{Tucker-Smith:2001myb}, we find that there is interesting parameter space which cannot be probed by direct-detection experiments but where neutrino experiments can provide complementary sensitivity. 

We find that the sensitivity of neutrino experiments depends on the flavor structure of inelastic dark matter interactions. We have considered separately couplings to both heavy-quarks and light-quarks in dark matter annihilation in the Sun. For coupling to heavy-quarks, both the spike and shoulder neutrinos are detectable, and we find that current limits from Super-Kamiokande \cite{Super-Kamiokande:2015xms} and IceCube \cite{IceCube:2021xzo} already rule out a large part of the inelastic dark matter parameter space that is inaccessible to direct-detection experiments (see \Cref{fig:main_sigma-delta} and \Cref{fig:main_sigma-m}). In this channel, our projected sensitivity for contained events in DUNE is weaker than the current limits from upward-going muons from Super-Kamiokande, and IceCube. We encourage Super-Kamiokande to perform a dedicated search for inelastic dark matter using their new atmospheric neutrino data \cite{Super-Kamiokande:2017yvm}. We also encourage IceCube and Hyper-Kamiokande to search for inelastic dark matter in the Sun. For coupling to light-quarks, only the spike is detectable as the flux of shoulder neutrinos is relatively small. In this scenario, DUNE excels due to its better performance at low-energy, and will probe currently unexplored parts of the inelastic dark matter parameter space. In this paper, we have extended the analysis of Ref. \cite{DUNE:2021gbm} to include a wider range of mass splittings. In the future, a positive signal from spike in DUNE, and null results from shoulder neutrinos in other neutrino experiments will strongly hint towards inelastic dark matter coupling to light-quarks only.

The DUNE collaboration could improve our analysis by including the effects of detector resolution on the reconstruction of neutrino energy and direction. For the spike neutrinos, the DUNE collaboration finds that these effects reduce the sensitivity by a factor of $\sim$10 \cite{DUNE:2021gbm}. The shoulder neutrinos are higher in energy, and we expect the detector resolution to be better, so these effects should be smaller. For shoulder neutrinos, it is reasonable to assume that the angular distribution of the background atmospheric neutrinos is isotropic. Including the location-specific  zenith-angle dependence of the atmospheric neutrinos in the Monte Carlo generators will result in a small changes in the projected sensitivity presented in this paper. We encourage the DUNE collaboration to consider a dedicated study of GeV-scale atmospheric neutrinos in the future.

\section*{Acknowledgements}
We would like to thank Sergio Palomares-Ruiz, Kate Scholberg, Raj Gandhi, Sandra Robles, Kajetan Niewczas, Gaurav Tomar, Ivan Martinez-Soler, and Juergen Reichenbacher for helpful discussions and comments. BC and MHR are supported by US Department of Energy Grant DE-SC-0010113. IS is supported by the U. S. Department of Energy Grant DE-FG02-13ER41976/sc-0009913. C. Rott acknowledges support from NSF Grant No. PHY-2309967 and from the National Research Foundation of Korea (NRF) for the Basic Science Research Program NRF-2020R1A2C3008356. For facilitating portions of this research, BC and MHR wish to acknowledge the Center for Theoretical Underground Physics and Related Areas (CETUP*), The Institute for Underground Science at Sanford Underground Research Facility (SURF), and the South Dakota Science and Technology Authority for hospitality and financial support, as well as for providing a stimulating environment. It must be noted that this analysis has been performed independently by the authors and does not reflect the opinions of the DUNE collaboration.

\bibliographystyle{JHEP}
\bibliography{ref.bib}

\end{document}